\documentclass[a4paper,11pt]{article}
\pdfoutput=1
\usepackage{jheppub}
\usepackage{graphicx}
\usepackage{subcaption}
\usepackage{dsfont}
\usepackage{xcolor}
\usepackage{slashed}
\usepackage{physics}
\title{Simple Perturbatively Traversable Wormholes from Bulk Fermions}

\author[a]{Donald Marolf}
\author[a]{and Sean McBride}

\affiliation[a]{Department of Physics, University of California, Santa Barbara, CA 93106, USA}

\emailAdd{marolf@physics.ucsb.edu}
\emailAdd{seanmcbride@physics.ucsb.edu}

\abstract{A new class of traversable wormholes was recently constructed which relies only on local bulk dynamics rather than an explicit coupling between distinct boundaries. Here we begin with a four-dimensional Weyl fermion field of any mass $m$  propagating on a classical background defined by a ${\mathds Z}_2$ quotient of (rotating) BTZ $\times \, S^1$. This setup allows one to compute the fermion stress-energy tensor exactly. For appropriate boundary conditions around a non-contractible curve, perturbative back-reaction at any $m$ renders the associated wormhole traversable and suggests it can become eternally traversable at the limit where the background becomes extremal. A key technical step is the proper formulation of the method of images for fermions in curved spacetime. We find the stress-energy of spinor fields to have important kinematic differences from that of scalar fields, typically causing the sign of the integrated null stress-energy (and thus in many cases the sign of the time delay/advance) to vary around the throat of the wormhole. Similar effects may arise for higher-spin fields.}

\begin{document}
\maketitle

\section{Introduction}

Wormholes have long been a source of fascination both in the scientific literature \cite{einstein, fuller, ellis, morristhorne, MTY} and in science fiction \cite{James:2015ima} as a potential tool for producing superluminal travel. In classical general relativity, wormholes are nontraversable due to constraints on causality from the null energy condition (NEC), which implies the topological censorship theorems \cite{Friedman:1993ty,Galloway:1999bp}. With the assumption of global hyperbolicity, these theorems require causal curves that start and end at the boundary to be deformable to the boundary in both asymptotically flat and asymptotically anti-de Sitter (AdS) contexts\footnote{Traversable wormholes can be constructed if one drops the requirement of global hyperbolicity, e.g. by introducing NUT charge \cite{Ayon-Beato:2015eca,PhysRevD.96.025021}.}. Adding quantum corrections allows fluctuations that violate the null energy condition, though the topological censorship theorems continue to apply in contexts where the integrated null energy along causal curves is non-negative so that the averaged null energy condition (ANEC) holds.

More generally, however, one might expect the ANEC to hold only along achronal curves; i.e., along only the fastest causal curves connecting two events \cite{Graham:2007va}. For example, in spacetimes with non-contractible closed spacelike curves, the Casmir effect causes the ANEC to fail. Indeed, it is this achronal ANEC (AANEC) that follows from the generalized second law \cite{Wall:2009wi}. Traversable wormholes are thus allowed, so long as it takes longer to travel through the wormhole than to go around. Similar conclusions follow from requiring boundary causality in the context of AdS/CFT \cite{Maldacena1997}.

Consistent with such expectations, traversable wormholes were constructed by Gao, Jafferis, and Wall (GJW) \cite{Gao:2016bin} by introducing a time-dependent coupling between the two otherwise disconnected asymptotically AdS$_3$ boundaries. This nonlocal coupling induces a bulk perturbation of a scalar quantum field whose back-reaction allows causal curves to run between the two boundaries. The averaged null energy along these trajectories is negative, but the nonlocal boundary coupling transmit signals from one boundary to the other more quickly than they could travel through the wormhole. In this sense the AANEC is satisfied. Similar setups have been studied for AdS$_2$ and dual SYK models \cite{Maldacena:2017axo, Mertens:2017mtv, Maldacena:2018lmt} and for rotating AdS$_3$ wormholes in \cite{Caceres2018}. Furthermore, since such traversable wormholes provide a holographic dual of certain quantum teleportation protocols, they are of broader interest in connection with the ER = EPR \cite{Maldacena:2013xja, Susskind:2017nto} and GR = QM \cite{Susskind:2017ney} conjectures.

Nonlocal boundary couplings of the sort used in GJW were expected to model more general backgrounds in which causal curves can travel from one wormhole mouth to the other; e.g., when both mouths are embedded in the same asymptotically flat or asymptotically AdS region of spacetime. This expectation was recently verified by two complementary constructions. In \cite{Maldacena:2018gjk}, a nearly-AdS$_2$ approximation was used to construct a time-independent asymptotically flat four-dimensional wormhole. This approach allowed \cite{Maldacena:2018gjk} to address many non-perturbative issues. The contrasting approach of \cite{Fu:2018oaq} (see also \cite{Fu2019}) used a perturbative framework to argue that standard local quantum fields on a broad class of such classical wormhole geometries
have Hartle-Hawking-like states whose stress-energy back-reacts on the geometry to renders the wormholes traversable.

The class of backgrounds $M$ considered by \cite{Fu:2018oaq} were of the form $M = \tilde M/{\mathds Z}_2$, where the covering space $\tilde M$ contains black holes with Killing horizons and well-defined Hartle-Hawking states for any quantum fields. This context includes many examples with familiar wormhole topologies having fundamental group $\pi_1 = {\mathds Z}$, though it also includes what one might call torsion wormholes with e.g. $\pi_1 = {\mathds Z}_2$. For linear quantum fields, the sign of the back-reaction, and thus whether or not it makes the wormhole traversable, is controlled by a choice of periodic or anti-periodic boundary conditions under the action of this $\mathds{Z}_2$. For scalar fields, for instance, it depends on whether the $\mathds{Z}_2$ isometry $J$ maps $\phi(x)$ to $\phi(Jx)$ or to $-\phi(Jx)$, and thus whether $\phi$ satisfies periodic or anti-periodic boundary conditions on the quotient spacetime $M$. Here we follow \cite{Fu:2018oaq} in using the term wormhole to refer to any setting where curves both starting and ending at the boundary cannot be smoothly deformed to the boundary. Such curves would classically be forbidden from being causal by the aforementioned topological censorship results of \cite{Friedman:1993ty,Galloway:1999bp}.

In certain asymptotically AdS$_3$ examples of such torsion wormholes, \cite{Fu:2018oaq} was able to exactly compute quantum stress-energy tensor expectation values for general free bulk scalar fields. The study of more general quantum fields is clearly of interest, though in curved spacetime computations involving higher spin fields can lead to significant technical complications. Here we take a first step in this direction by computing the quantum back-reaction from Weyl fermions of any mass $m$ in the torsion wormholes of \cite{Fu:2018oaq}. Again, we find that stress-energy tensor expectation values can be computed exactly.

The outline of the paper is as follows. In section \ref{sec:fgm}, we review the construction of $\mathds{Z}_2$ wormholes from \cite{Fu:2018oaq} and find geometries useful for our consideration of spinor fields. In the relevant case the spacetime $M$ is a ${\mathds Z}_2$ quotient of the rotating Ba\~nados-Teitelboim-Zanelli black hole \cite{Banados:1992wn,Banados:1992gq} (rBTZ) times $S^1$, so that  $M = ({\rm rBTZ} \times \; S^1)/{\mathds Z}_2$. We flesh out the details of this case in section \ref{sec:spinors}, providing an analytic expression for the null stress-energy tensor on the horizon of $M$ for a spinor field of arbitrary mass.  The above quotient breaks rotational symmetry, and we find the sign of the integrated null energy to generally depend on the BTZ angular coordinate $\phi$.  But the average is non-zero when the black hole rotates, and is negative with the appropriate choice of periodicity. Following \cite{Fu:2018oaq} we compute $T \expval{ \Delta V}$, where $T$ is the black hole temperature and $ -\expval{ \Delta V}$ measures the expectation value of the time-advance governing traversability of the wormhole.   As in the scalar case, we find $T \expval{ \Delta V}$  to be independent of $T$ for all fermion masses, suggesting that $\expval{ \Delta V}$ diverges  as $T \rightarrow 0$ and that bulk spinors alone would suffice to yield an eternally traversable wormhole in that limit. We conclude in section \ref{sec:conclude} by discussing the extension of our results to other higher spin particles which, like the spinor, have exactly soluble propagators in AdS$_d$. Working our way all the way up to spin-2 would allow understanding of the effect of linearized gravitons on such wormholes; this remains a goal for future work.

\section{Preliminaries and Review}
\label{sec:fgm}

We begin in section \ref{sec:spacetimeRev} with a brief review of certain asymptotically AdS (or asymptotically AdS $\times \, X$) ${\mathds Z}_2$-wormholes studied in \cite{Fu:2018oaq}. We then recall in section \ref{sec:sTkk} how the stress-energy tensor of quantum scalar fields can be computed exactly in these backgrounds. The extension of such computations to fermions is outlined in section \ref{sec:fermi3}, which in particular describes further properties required for fermions to yield non-trivial results. The example from \cite{Fu:2018oaq} satisfying these properties is called the Kaluza-Klein zero-brane orbifold (KKZBO) spacetime and is discussed in detail in section \ref{subsec:KKZBO}.

\subsection{Review of AdS $\mathds{Z}_2$-Wormholes}
\label{sec:spacetimeRev}

The exactly solvable models of \cite{Fu:2018oaq} involved quantum fields in a Hartle-Hawking-like state propagating on $\mathds{Z}_2$ quotients $M = \tilde M/\mathds{Z}_2$ of BTZ and BTZ $\times \; S^1$.  This setting is useful as the Killing symmetry of BTZ preserves the BTZ Hartle-Hawking state. As a result, in that state on the covering space $\tilde M$ the symmetry requires the expected null-null component of the stress-energy tensor $\expval{T_{kk}}_{\tilde M}$ to vanish on the BTZ horizon for any quantum field. This symmetry is then broken by the $\mathds{Z}_2$ quotient, so on the physical spacetime $M$ the horizon expectation value $\expval{T_{kk}}_{M}$ can be non-zero. Nevertheless, the simplicity of BTZ can be used to provide an analytic expression for $\expval{T_{kk}}_{M}$. The integrated $\expval{T_{kk}}_{M}$ on the horizon can then be evaluated numerically and combined with first-order perturbation theory to compute the back-reaction. We begin by reviewing simple examples of the above geometries and techniques. In this section we set the angular momentum of the black hole to zero and consider only non-rotating BTZ.

Without rotation, the BTZ metric can be written in the form
\begin{equation}
\label{eq:BTZds}
ds^2 = \frac{1}{(1+UV)^2} \left( -4 \ell^2 dU dV + r_+^2(1-UV)^2 d\phi^2 \right).
\end{equation}
Here $\ell$ is the AdS length scale (which we will often take to be one) and $2\pi r_+$ is the length of the BTZ horizon. We use Kruskal-like coordinates $(U, V, \phi)$ where $U$ and $V$ parameterize the null directions.  The metric \eqref{eq:BTZds} is global $\textrm{AdS}_3$ when $\phi$ takes values in $(-\infty, \infty)$, but gives the global BTZ black hole when one makes the identification $\phi \sim \phi + 2 \pi$ \cite{Banados:1992gq}.

A variety of interesting spacetimes can be constructed as quotients of \eqref{eq:BTZds} or of tensor products with some other simple factor $X$. For example, the $\mathds{R}P^2$ geon \cite{Louko:1998hc} is a quotient of \eqref{eq:BTZds} under the isometry $J_1: (U,V, \phi) \mapsto (V,U, \phi + \pi)$; see figure \ref{fig:rp2} below. Note that $J_1^2$ is the identity and that the quotient BTZ$/J_1$ contains a non-contractible cycle with ${\mathds Z}_2$ homotopy that is not deformable to the boundary and can be represented by the closed curve $\phi \in [0,\pi]$ for any $U(\phi)=V(\phi)$. Another interesting related spacetime is the Kaluza-Klein end-of-the-world brane geometry (hereafter KKEOW) $({\rm BTZ} \times \; S^1)/J_2$, where the isometry $J_2$ now acts on the angle $\theta$ associated with the internal $S^1$ as well as on the BTZ factor. In particular, $J_2: (U, V, \phi, \theta) \mapsto (V, U, \phi, \theta + \pi)$. Here the quotient is smooth but Kaluza-Klein reduction on $S^1$ gives a singular spacetime with what may be called an end-of-the-world brane at $U=V$. Again, $J_2^2$ is the identity and the quotient contains a non-contractible cycle with ${\mathds Z}_2$ homotopy that is not deformable to the boundary and can be represented by the closed curve $\theta \in [0,\pi]$, $\phi = \textrm{constant}$ for any $U(\theta,\phi)=V(\theta,\phi)$.
\begin{figure}
\centering
\includegraphics[width = 8cm]{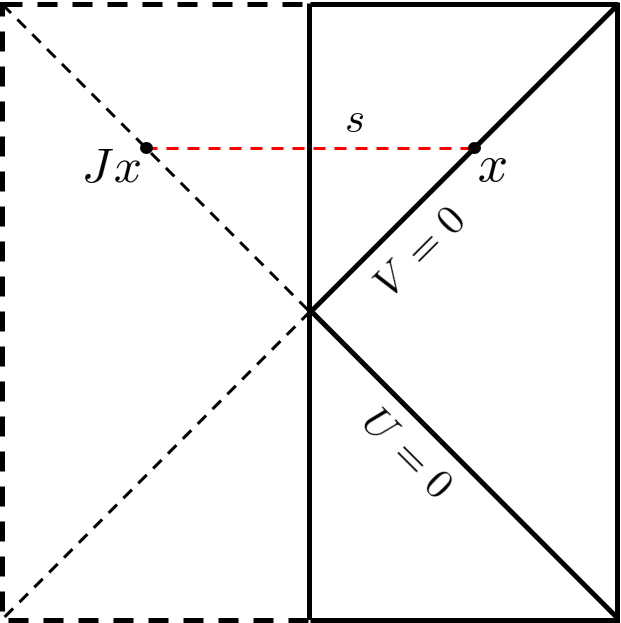}
\caption{Conformal diagram of the $\mathds{R}P^2$ geon, KKEOW, and KKZBO geometries, with null coordinates $(U, V)$ indicated. The left region (dashed lines) is the image of the right (solid lines) under the $\mathds{Z}_2$ isometry $J$ which exchanges $U \leftrightarrow V$, so that the right alone may be used to represent the quotient.  The desired isometries $J$ also act on
$S^1$ factors (just $S^1$ for the $\mathds{R}P^2$ geon and $S^1 \times S^1$ for the KKEOW and KKZBO geometries) not shown in the figure.   The geodesic proper distance $s$ between a point $x$ on the horizon and its image $Jx$ is non-zero and spacelike in all cases, though for $U=V=0$ this is a result of the additional action of $J$ on suppressed angles $\phi$ and/or $\theta$ not shown in this figure. }
\label{fig:rp2}
\end{figure}

Both of the above quotients are non-orientable, but an orientable spacetime $({\rm BTZ} \times \; S^1)/J_3$ can be obtained by allowing the isometry $J_3$ to act on both $\theta$ and $\phi$.  In particular, \cite{Fu:2018oaq} studied the Kaluza-Klein zero-brane orbifold (KKZBO) given by
$J_3: (U, V, \phi, \theta) \mapsto (V, U, -\phi, \theta + \pi)$.  As before, the quotient is smooth but singularities arise when it is Kaluza-Klein reduced on the $S^1$.  These singularities are localized at $U=V$ with $\phi =0,\pi$ and may be called zero-brane orbifolds as they represent point-like defects in the resulting three-dimensional spacetime.  Once again, $J_3^2$ is the identity and the quotient contains a non-contractible cycle with ${\mathds Z}_2$ homotopy that is not deformable to the boundary and is now associated with the zero branes. It can be represented by the closed curves $\theta \in [0,\pi]$ at $\phi =0$ or $\pi$ for any $U(\theta)=V(\theta)$.  The KKZBO is more complicated than the above examples as it breaks translational symmetry in $\phi$.  However, as noted in \cite{Fu:2018oaq}, it has the advantage of extending to rotating black holes where both $J_1$ and $J_2$ cease to be isometries.  In particular, the KKZBO admits an interesting extreme limit in which back-reaction from quantum fields can become large. As reviewed in section \ref{sec:fermi3} below, even without rotation, the fact that the KKZBO is orientable makes it a much more interesting context in which to study fermions.

We are interested in finding the quantity $\expval{\Delta V}$, the expected null time delay of a geodesic starting at $U = -\infty$ and ending at $U = \infty$ induced by back-reaction from quantum fields.  A negative value $\expval{\Delta V} < 0$ indicates that the wormhole becomes traversable, as a null geodesic fired from the left boundary in the distant past then arrives at the right boundary at a finite future time with (on average) coordinate $V=\expval{\Delta V}$. As shown in \cite{Gao:2016bin}, in the linearized approximation and in the presence of rotational symmetry, $\expval{\Delta V}$ in non-rotating BTZ (or BTZ $\times X$) is related to the integrated null stress-energy tensor along the horizon through
\begin{equation}
\label{eq:DV}
\expval{\Delta V} = 4 \pi G_N \int_0^\infty dU \expval{T_{kk}}.
\end{equation}
Here $T_{kk} \equiv T_{\mu \nu} k^\mu k^\nu$ for a null horizon generator $k^\mu \partial_\mu = \partial_U$. The extension to rotating BTZ and to broken rotational symmetry will be reviewed below in section \ref{subsec:KKZBO} below.

As noted in the introduction, symmetry requires
$\expval{T_{kk}}$ to vanish in the Hartle-Hawking state on BTZ or ${\rm BTZ} \times X$, but we will in fact compute $\expval{T_{kk}}$ in the quotient $({\rm BTZ} \times X)/{\mathds Z}_2$ where the symmetry is broken.  At linear order in perturbation theory we may then lift the resulting $\expval{T_{kk}}$ back to the $({\rm BTZ} \times X)$ covering space, compute back-reaction on the metric using  \eqref{eq:DV}, and take the quotient of the result to obtain a self-consistent semi-classical solution.

The stress-energy tensor of a scalar field $\phi$ takes the form
\begin{equation}
T_{\mu \nu} = \partial_\mu \phi \partial_\nu \phi - \frac{1}{2}g_{\mu \nu}g^{\rho \sigma} \partial_\rho \phi \partial_\sigma \phi  - \frac{1}{2}g_{\mu \nu}m^2 \phi^2.
\label{eq:Tmngen}
\end{equation}
Contracting with the null vectors eliminates all but the first term, so $\expval{T_{kk}}$ is a coincident limit of derivatives acting on the two-point function $\expval{\phi \phi}$.  The pointwise contribution to the null stress-energy tensor can thus be determined by derivatives of the scalar Green's function. Since $k^\mu \partial_\mu = \partial_U$, we may write
\begin{equation}
\expval{T_{kk} (x)} = \lim_{x \rightarrow x'} \expval{\partial_U \phi (x) \partial_{U} \phi (x')}
\end{equation}
where the computation is to be done in the Hartle-Hawking state on the quotient geometry $M$.

As reviewed in \cite{Fu:2018oaq}, for linear quantum fields correlations functions in this state may be computed using the method of images. In particular, one may identify the quantum field $\phi(x)$ on the quotient $M$ with an appropriately scaled linear combination of the scalar field $\tilde \phi (x)$ on the covering space $\tilde M$ and its image under the relevant isometry $J$:
\begin{equation}
\phi (x) \equiv \frac{1}{\sqrt{2}} \left( \tilde \phi (x) \pm \tilde \phi (J x) \right).
\label{HHstate}
\end{equation}
The choice of sign in \eqref{HHstate} corresponds to a choice of periodic ($+$) or anti-periodic ($-$) boundary conditions on the $\mathds{Z}_2$ cycle of the quotient.

The stress-energy tensor
\eqref{eq:Tmngen} is thus a sum of four terms involving coincident limits of two point functions involving all possible pairings of $\tilde \phi(x)$ and $\tilde \phi(Jx)$. Terms involving coincident points in the covering space are proportional to the stress-energy tensor in the Hartle-Hawking state on $\tilde M$ and thus vanish by the symmetry noted above. Thus all potentially divergent terms vanish and $\expval {T_{kk}}_M$ is explicitly finite. Non-zero contributions come only from the two cross terms. These contribute equally and give
\begin{equation}
\expval{T_{kk} (x)}_M = \pm \expval{\partial_U \phi (x) \partial_{U} \phi (J x)}_{\tilde M}.
\end{equation}
We can now see that creating a traversable wormhole is simple, as unless some coincidence or symmetry imposes $\expval{T_{kk}}_M =0$, one will find $\expval{T_{kk}}_M$ to be negative for one choice of periodic or anti-periodic scalars.  As verified by detailed computations in \cite{Fu:2018oaq}, the correct choice turns out to be periodic boundary conditions for the ${\mathds RP}^2$ geon, the KKEOW spacetime, and the KKZBO geometry.

\subsection{Explicit Calculation}
\label{sec:sTkk}
We now review further explicit scalar results from \cite{Fu:2018oaq}. The scalar two-point function on empty AdS$_3$ is known exactly \cite{Ichinose:1994rg}.  It may be written in the form
\begin{equation}
G_{\textrm{AdS}_3} (Z) = \frac{1}{4 \pi} \left( Z^2 - 1 \right)^{-1/2} \left( Z + \left( Z^2 - 1 \right)^{1/2} \right)^{1 - \Delta},
\end{equation}
where $Z \equiv 1 + \sigma (x, x')/\ell^2$, $\sigma (x, x')$ is the half squared-geodesic distance in an associated four dimensional embedding space $\mathds{R}^{2,2}$, and $\Delta$ is given by the same formula as the scaling dimension of the associated operator in any dual CFT \cite{Aharony:1999ti},
\begin{equation}
\Delta = 1 \pm \sqrt{1 + m^2 \ell^2}
\label{eq:cft}
\end{equation}
where the $\pm$ denotes a choice of boundary conditions at AdS infinity. We will always take the $(+)$ boundary condition below, as this choice is always consistent with unitarity.  However, the $(-)$ choice can also be of interest; see \cite{Fu:2018oaq} for associated results in the context of the above ${\mathds Z}_2$ wormholes.

From \cite{Fu:2018oaq}, the half-squared geodesic distance in $\mathds{R}^{2,2}$ in our Kruskal-like coordinates is
\begin{align}
\sigma (x, x') & = \frac{\ell^2}{(1 +UV) (1+U' V')} \left[ (UV-1)(U'V'-1) \cosh \left( \frac{r_+ \left( \phi  - \phi' \right)}{\ell} \right) \right. \nonumber \\
&- \left. (1 +UV)(1 + U'V') + 2 (UV' +VU') \right].
\end{align}
We can use the fact that BTZ is a quotient space of AdS$_3$ to recast the BTZ Green's function as a sum over images in the AdS$_3$ covering space such that
\begin{equation}
\label{eq:BTZimages}
G_{\textrm{BTZ}} (Z) = \sum_{n \in \mathds{Z}} G_{\textrm{AdS}_3} \left( Z_n \right)
\end{equation}
where $Z_n \equiv 1 + \sigma (x, x'_n)/\ell^2$ and $\phi_n' = \phi' + 2 \pi n$. The additional isometry to produce the $\mathds{R}P^2$ geon means that we are concerned with cases where $\phi' = \phi + \pi$, so the sum over images is a sum over $\phi' = \phi + (2n + 1) \pi$ with $U' = V = 0$ and $V' = U$.

By substituting the expression for the embedding space geodesic distance, we can calculate $G_{\textrm{AdS}_3}$, and by extension the two-point function on the $\mathds{R}P^2$ geon. The full expression for $\expval{T_{kk}}$ is quite complicated, so we recall from \cite{Fu:2018oaq} only the result in the limit $r_+/\ell \rightarrow 0$.  Using $\expval{T_{kk}}_n$ to denote the contribution to $\expval{T_{kk}}$ from the $n$th term in the sum \eqref{eq:BTZimages}, for periodic scalars one finds in this limit the $n$-independent result
\begin{equation}
\begin{aligned}
\expval{T_{kk}}_n = \frac{\left( 1 + 2U (U + \sqrt{1 +U^2})\right)^{1 - \Delta}}{32 \pi U^3 (1 + U^2)^{5/2}}& \left[  1 + \left( 2U \sqrt{1 + U^2} + 8U^3 \sqrt{1 +U^2} \right) (\Delta - 1) + \right.
\\ & \left. 4U^4 (2 - 2\Delta + \Delta^2) + U^2 (6 - 8 \Delta + 4 \Delta^2)  \right].
\label{eq:Tkkboson}
\end{aligned}
\end{equation}
While the sum of \eqref{eq:Tkkboson} over $n$ clearly diverges, the result \eqref{eq:Tkkboson} will still prove useful in section \ref{sec:spinors}.   For general $r_+$, numerically integrating the full expression for periodic scalars yields a strictly negative result for each $\expval{T_{kk}}_n$, so the back-reaction does indeed render the wormhole traversable.

The calculation for the KKEOW geometry is slightly more involved.  For any given four-dimensional mass $m$, Kaluza-Klein (KK) reduction on the $S^1$ produces a tower of massive three-dimensional fields with effective masses given by
\begin{equation}
m_{eff}\ell = \sqrt{m^2 \ell^2 + \left( \frac{\ell}{R_{S^1}} \right)^2 p^2},
\label{eq:KKmass}
\end{equation}
where $p \in \mathds{Z}$ is the mode number on the internal $S^1$ with radius $R_{S^1}$. Since the three-dimensional two-point function \eqref{eq:BTZimages} is known exactly, it is useful to write the full four-dimensional stress-energy tensor as a sum over $p$ of contributions $\expval{T_{kk}}_{n,p}$ from the $n^{\textrm{th}}$ term in \eqref{eq:BTZimages} and the $p^{\textrm{th}}$ mode on the $S^1$.

Such contributions include a factor of
of $e^{i p \pi} = (-1)^p$ from the action of the isometry on $\theta$. These alternating signs play an important role, as the $n=0$ contributions $\expval{T_{kk}}_{n=0,p}$ all have non-integrable singularities at $U=0$.  Indeed, such terms are independent of $r_+$ and give
\begin{equation}
\label{eq:scalarsing}
\expval{T_{kk}}_{n=0,p} = \frac{1}{32 \pi U^3} + \frac{3 - 8 \Delta + 4 \Delta^2}{64 \pi U} + \frac{-2 \Delta + 3 \Delta^2 - \Delta^3}{6 \pi} +O(U).
\end{equation}
Since the four-dimensional state is Hadamard and the quotient is smooth, the full stress-energy tensor can have only integrable singularities and the non-integrable terms in \eqref{eq:scalarsing} must cancel when summed over $p$. For numerical calculations, we can simply choose some large $N$ and sum over modes with $|p| \le N$ if we also impose a cutoff at small $U$ to avoid possible issues from incomplete cancellations of such terms at finite $N$; see \cite{Fu:2018oaq} for details\footnote{It was also shown in \cite{Fu:2018oaq} that the explicit terms above vanish when summed over $p$ using Dirichlet $\eta$ regularization for the $1/U^3$ and $1/U$ terms and Abel summation for the constant term. One may thus rewrite $\expval{T_{kk}}$ as a sum of explicitly finite terms.}. Again, the integrated $\expval{T_{kk}}$ is negative for all periodic scalars of the type discussed above.

The KKZBO computations are similar but the integrated $\expval{T_{kk}}$ now depends on $\phi$.  For $\phi\neq 0,\pi$ each term $\expval{T_{kk}}_{n,p}$ is finite and continuous.  However, at $\phi=0,\pi$ the action of $J_3$ coincides with the KKEOW actions of $J_3$.  As a result, the expressions for $\expval{T_{kk}}_{n,p}$ coincide there as well.  In particular, corresponding care is required for the $n=0$ modes. For the simple scalars discussed here, periodic boundary conditions make the integrated $\expval{T_{kk}}$ negative along all horizon generators, though \cite{Fu:2018oaq} also found more complicated examples where the sign of $\expval{\Delta V}$ varies with $\phi$.

\subsection{Subtleties of Spinor Representations}
\label{sec:fermi3}

We now consider how the above constructions should be generalized to accomodate bulk fermion fields. To begin, recall that spin groups admit two inequivalent fundamental representations in odd dimensional spacetimes.  One may think of this choice as arising from two distinct possible definitions of $\gamma^{d-1} = \pm i^{-\frac{d-3}{2}} \prod_{i=0}^{d-2} \gamma^i$, both of which form a representation of the Clifford algebra $\{\gamma^a, \gamma^b\} = 2 \eta^{a b}$ \cite{Polchinski:1998rr}, or as choice of
\begin{equation}
\label{eq:3dreps}
\prod_{i=0}^{d-1} \gamma^i = \pm \mathds{I}_{2^{(d-1)/2}}.
\end{equation}
We shall simply denote the two choices as $\gamma_A^a$ and $\gamma_B^a$ below, with the understanding that the sign on the right-hand-side of \ref{eq:3dreps} is $+$ in the $A$ representation and $-$ in the $B$ representation, with $\gamma_A^a = - \gamma_B^a$.

Since the two spinor representations differ by the sign of all $\gamma^{\mu}$, they are related by the action of the three-dimensional parity operator, which we may take to be any orientation-reversing isometry of 3D Minkowski space. As a result, on a non-orientable spacetime like the ${\mathds RP}^2$ geon, spinors can be well-defined only when both representations are present.  In particular, our theory on the geon must contains two spinor fields $\psi_A$ and $\psi_B$, each corresponding to a different representation, but which are exchanged when one traverses a non-contractible curve that reverses orientation. While it is inconsistent to have a single fermion corresponding to either representation alone, this AB doublet of fermions yields a well-defined theory. In this context, we may use the Lagrangian \cite{deJesusAnguianoGalicia:2005ta}
\begin{equation}
\label{eq:dJAG}
\mathcal{L} = \det(e)\left[\overline{\psi_A} ( i \gamma_A^\mu D_\mu - m) \psi_A + \overline{\psi_B} ( i \gamma_B^\mu D_\mu  - m) \psi_B\right]
\end{equation}
where $\det (e)$ is the vielbein determinant and $\gamma^\mu_{A,B}$ denote the gamma matrices for each fermion representation.  In terms of a method-of-images construction paralleling that for scalars in \cite{Fu:2018oaq} in the scalar case, the Lagrangian on the BTZ covering space must also take the form \eqref{eq:dJAG}.

Let us now consider the analogue of the scalar relation \eqref{HHstate}.  For fields of non-zero spin, adding the field operators at distinct points $x$ and $Jx$ requires one to first find some way to identify the corresponding two tangent spaces.  It is natural to use the isometry $J$ and, as discussed in section \ref{sec:spinors} below, any isometry can be extended to a map $\hat J$ taking spinors at $x$ to spinors at $Jx$.  In particular, the spinor field $\hat J \tilde \psi$ evaluated at $x$ is an operator built from $\tilde \psi(Jx)$. We may thus write
\begin{equation}
\label{eq:spinimage}
\psi (x) \equiv \frac{1}{\sqrt{2}} \left(\tilde  \psi(x) +  \left[j \tilde \psi\right] \right) ( x ).
\end{equation}
Since the equation of motion is linear, it is preserved by $\hat J$ if $\hat J$ preserves the vielbein and the $\gamma^a$. We will use these conditions in section \ref{subsec:KKZBO} to define the appropriate extension $\hat J$ of the isometry $J$. The ansatz \eqref{eq:spinimage} will then satisfy the equation of motion on the quotient $M$ when $\tilde \psi$ satisfies the corresponding equation on the covering space $\tilde M$.  One may also check that canonical normalization of $\tilde \psi$ gives canonical normalization of $\psi$.  Thus \eqref{eq:spinimage} is the desired method-of-images ansatz. As for scalars, correlation functions in spinor Hartle-Hawking states on $M$ and $\tilde M$ will again be related by \eqref{eq:spinimage}.

We now consider the implications for stress-energy tensors on the ${\mathds RP}^2$ geon, which we represent as BTZ$/J_1$. The important point here is that $J_1$ reverses orientation and must thus interchange $\psi_A$ and $\psi_B$.  In other words, in this context we may write \eqref{eq:spinimage} more explicitly as
\begin{align}
\label{eq:spinimageAB}
\psi_A (x) \equiv \frac{1}{\sqrt{2}} \left(\tilde  \psi_A(x) +  \left[\hat  J_1 \tilde \psi_B\right] (x ) \right), \ \ \
\psi_B (x) \equiv \frac{1}{\sqrt{2}} \left(\tilde  \psi_B(x) +  \left[\hat J_1 \tilde \psi_A  \right] (x ) \right),
\end{align}
where $\left[ J_1 \tilde \psi_B\right] (x )$ is built from the operator $\tilde \psi_B (J_1 x)$.
Since the Lagrangian \eqref{eq:dJAG} contains no interactions, the full stress-energy tensor is of course a sum of separately conserved stress-energy tensors for the $A$ and $B$ fields.  Let us consider first the expectation value of the $A$ stress-energy tensor, which is a quadratic composite operator much like the stress-energy tensor of a scalar field.  As in the scalar case, using the first line in \eqref{eq:spinimageAB} generates four terms. The term involving coincident points at $x$ gives the $A$ stress-energy tensor on the covering space which vanishes by the same symmetry described above for scalars, and the term involving coincident points at $J_1x$ gives the similarly-vanishing $B$ stress-energy tensor. The final two terms are cross-terms built from the correlators $\expval{ \psi_A (x) \psi_B(J_1x)}$ and $\expval{ \psi_B (J_1x) \psi_A(x)}$ in the Hartle-Hawking state on the covering space $\tilde M$. The the lack of an interaction term in \eqref{eq:dJAG} means that such $AB$ cross-correlators vanish identically.  So despite the breaking of the Killing symmetry on the ${\mathds RP}^2$ geon, the $A$ stress-energy tensor continues to vanish on the horizon, as does the $B$ stress-energy tensor by the same argument. The shift $\Delta V$ thus remains zero and first-order back-reaction does not render the wormhole traversable.

One might ask if fermionic back-reaction is more interesting on the KKEOW spacetime.  There the full spacetime is four-dimensional, so there is only one representation of the Clifford algebra and it is invariant under orientation-reversal. Thus we need only consider a single Weyl fermion. However, as for the scalar case one may proceed by dimensional reduction to three dimensions, where one obtains a set of uncoupled free fields that come in pairs like the fields $AB$ discussed above. In particular, the $AB$ fields in each doublet are related by reversal of orientation on the three-dimensional (orbifold) base space. In terms of each pair, the discussion proceeds precisely as for the $\mathds{R}P^2$ geon above, and the full stress-energy tensor again vanishes. From the  four dimensional perspective the point is that even-dimensional spinors admit a conserved notion of chirality, and that the two chiralities are exchanged by the orientation-reversing isometry $J_2$. The cross-terms in the stress-energy tensor are then built from two-point functions between Weyl spinor components of opposite chirality, which vanish in  Hartle-Hawking state.

The upshot of this discussion is that in any ${\mathds Z}_2$ quotient of a spacetime with a Killing symmetry (in any spacetime dimension) the contribution of spinors to expected stress-energy tensors on the horizon will vanish unless the quotient is orientable. For this reason we focus on the orientable KKZBO spacetime in the remainder of this work.  The generalization of our discussion thus far to this spacetime is reviewed in section \ref{subsec:KKZBO} below.

\subsection{The KKZBO Spacetime and Back-reaction}
\label{subsec:KKZBO}
As mentioned in section \ref{sec:spacetimeRev}, the KKZBO spacetime can be defined in the presence of rotation.  In particular, it admits an extreme limit where \cite{Fu:2018oaq} found that back-reaction from matter sources becomes large.  Perturbations that render the wormhole traversable may thus naturally create an eternally traversable wormhole in this limit in agreement with \cite{Maldacena:2018lmt,Maldacena:2018gjk}. We briefly review these results here for later use in studying back-reaction from the stress-energy tensor of our Weyl fermions.

To begin, recall that the rotating BTZ (${\rm rBTZ}$) metric in our Kruskal-like coordinates $ (U,V,\phi) $ takes the form
\begin{equation}
ds^2 = \frac{1}{(1+UV)^2} ( -4\ell^2 dU dV + 4\ell r_{-}(UdV - VdU) d \phi + \left[ r_+^2(1-UV)^2 +4UVr_{-}^2 \right] d\phi^2 ).
\label{eq:metric}
\end{equation}
We are interested in the Cartesian product of \eqref{eq:metric} with an $S^1$ of radius $R_{S^1}$, and thus with line element $R_{S^1}^2 d\theta^2$. Defining
\begin{equation}
\label{eq:J3}
J_3: (U, V, \phi, \theta) \rightarrow (V, U, -\phi, \theta + \pi)
\end{equation}
as in section \ref{sec:spacetimeRev}, it is clear that the $(U,V,\phi)$ part of $J_3$ preserves \eqref{eq:metric} and the $\theta$ part preserves the metric on $S^1$.

To understand the implications of the $\expval{T_{kk}}_{\textrm{KKZBO}}$ that we will compute in section \ref{sec:spinors}, we must understand the back-reactions of such a source on the geometry. As in the case without spin, one may lift the source $\expval{T_{kk}}_M$ computed on the quotient $M=\textrm{KKZBO}$ to the covering space $\tilde M =\textrm{rBTZ} \times S^1$, compute the associated back-reaction on $\tilde M = {\rm rBTZ} \times S^1$, and then quotient the result again by $J_3$. This lift and quotient procedure gives the same result as computing back-reaction directly on the KKZBO spacetime.  And since we preserve rotational symmetry on the internal $S^1$, the problem can be Kaluza-Klein reduced to studying back-reaction on three-dimensional $\rm{rBTZ}$. Indeed, section \ref{sec:spinors} will directly compute the effective three-dimensional stress-energy $\expval{T_{kk}}^{\textrm{3d}}_{\textrm{KKZBO}}$, which is just the integral over the internal $S^1$ of $\expval{T_{kk}}_{\textrm{KKZBO}}$.

We thus require the generalization of \eqref{eq:DV} to rotating BTZ and to sources that break rotational symmetry. As shown in \cite{Fu:2018oaq}, the correct result is
\begin{equation}
\label{eq:DVrot}
\Delta V = \frac{1}{4\ell^2} \int_0^{\infty} dU h_{kk} = \frac{2\pi G_N}{\ell^2}  \int_{-\infty}^{\infty} \int_{-\pi}^{\pi} d\phi' dU H (\phi - \phi') \expval{T_{kk}}^{3d}(\phi'),
\end{equation}
where the Green's function $H (\phi - \phi')$ is usefully described as a sum over Fourier modes of the form
\begin{equation}
\label{eq:HGF}
H (\phi - \phi' ) =\sum_q e^{iq (\phi-\phi')} H_q, \ \ \ H_q = \frac{1}{2 \pi} \frac{2\ell^2 r_+^2}{r_+^2-r_-^2 -2iqr_- + \ell^2 q^2}.
\end{equation}

Perhaps the most interesting feature of \eqref{eq:HGF} is that the zero-mode $H_{q=0} = \frac{\ell^2 r_+^2}{\pi \left( r_+^2-r_-^2 \right)}$, diverges in the extreme limit $r_+ \rightarrow r_-$.  As noted in \cite{Fu:2018oaq}, this invalidates our perturbative analysis for $r_+$ very close to $r_-$,  but it also suggests that a full non-perturbative analysis could produce a static, eternally traversable wormhole. We will thus be most interested in cases with $r_+ - r_-$ smaller than any classical scale, but with $r_+$ still far enough from $r_-$ that our perturbative treatment remains valid. Variations with $\phi$ are then a subleading effect as the $q\neq 0$ modes in \eqref{eq:HGF} remain finite at $r_+ = r_-$.  It is thus useful to focus on the average of the time delay \eqref{eq:DVrot} over $\phi$.  Since the temperature of rotating BTZ is
\begin{equation}
T = \frac{r_+^2-r_-^2}{2 \pi r_+ \ell^2} = \frac{2 r_+}{H_0},
\end{equation}
we may average $\langle \Delta V \rangle$ over the $\phi$-circle to write
\begin{equation}
\label{eq:DVav}
T \langle \Delta V \rangle_{\textrm{average}} = \frac{4G_N r_+}{\ell^2} \int_0^\infty \int_{-\pi/2}^{\pi/2} d \phi dU  \expval{T_{kk}}^{3d},
\end{equation}
where the factor of $4$ is associated with making use of symemtries to change the limits of integration relative to those in \eqref{eq:DVrot}. In the scalar case, the numerics in \cite{Fu:2018oaq} found the extreme limit of \eqref{eq:DVav} to be approximately independent of $r_+$.   We will find similar behavior below.

\section{Spinors and stress-energy tensors on the KKZBO spacetime}
\label{sec:spinors}

We now turn to the fermion details for stress-energy tensors on the KKZBO.  We begin in section \ref{subsec:viel} below by extending the KKZBO isometry $J_3$ of $\rm{rBTZ} \times S^1$ to act on spinor fields.  Section \ref{subsec:Tkk} then sets up the calculation of the desired stress-energy tensor components and section \ref{subsec:lim} studies a simplifying limit in preparation for more complete numerical calculations in section \ref{sec:numerics}.

\subsection{Spinor Extensions of the $J_3$ Isometry}
\label{subsec:viel}

Any isometry has a natural action on tensor fields via the associated diffeomorphism.  But spinor fields are not tensors, and are typically defined by attaching an internal tangent space to each point in spacetime.  This is done by choosing a vielbein $e^a_\mu$, which has a spacetime index $\mu = (U,V,\phi, \theta)$ and an internal index $a= (0,1,2,3)$.  Choosing the metric on the internal space to be $\eta_{ab} = \rm{diag}(-1,1,1,1)$, one may use any vielbein that satisfies $e_\mu^a e_\nu^b \eta_{ab} = g_{\mu \nu}$.  Any two such vielbeins are related by an internal $O(3,1)$ gauge transformation. For ${\rm rBTZ}\times S^1$ we take
\begin{equation}
\label{eq:eamu}
e^a_\mu = \frac{1}{1+UV}\begin{pmatrix} \ell&\ell&0&0\\\ell&-\ell&0&0\\- r_- (U-V)&-r_-(U+V)&r_+ (1-UV)&0\\0&0&0&R_{S^1}\left(1+UV\right)\end{pmatrix},
\end{equation}
where $a$ labels the columns and $\mu$ labels the rows.

The natural action of the isometry $J_3$ on $e^a_\mu$ is given by treating $e^a_\mu$ as a spacetime vector field for each $a$. Thus $J_3$ acts on spacetime indices $\mu$ and the point at which the field is evaluated but has no further action on the internal index $a$. As a result, in addition to acting on the arguments $(U,V, \phi, \theta)$ in \eqref{eq:eamu}, it also exchanges the $U$ and $V$ rows and changes the sign of all entries in the $\phi$ row. We note that this combined operation is {\it not} a symmetry of the vielbein.

However, as noted in section \ref{sec:fermi3}, it would be more useful for our method-of-images construction to have an operation $\hat J_3$ that leaves the vielbein invariant. Since the isometry $J_3$ preserves the metric, the vielbeins $J_3 e^a_\mu$ and $e^a_\mu$ can differ only by an internal $O(3,1)$ transformation. Choosing the right such transformation $j$ (such that $(je)^a_\mu = j^a_b e^b_\mu$) thus allows us to define a vielbein-preserving $\hat J_3 = J_3 \circ j$ as desired.  It is easy to check that this is the case for
\begin{equation}
\label{eq:intj}
j_a^b = \begin{pmatrix} 1&&& \\ &-1&& \\&&-1&\\&&&1\end{pmatrix}
\end{equation}
defines a $\hat J_3$ that preserves the vielbein \eqref{eq:eamu}.

To describe the corresponding action on spinors, one should view \eqref{eq:intj} as the action of an $O(3,1)$ transformation on covectors. There is then a corresponding action $\tilde j$ of this transformation on spinors, up to a sign to be discussed below associated with $O(3,1)$ being the double cover of the associated spin group, defined by requiring that $j$ leave invariant the four-dimensional Clifford algebra $\{\Gamma^a \}$ that satisfies $\{\Gamma^a, \Gamma^b\} = 2 \eta^{ab}$ and defines the four-dimensional spinor representation. The spinor-space matrix $\tilde j$ must satisfy
\begin{equation}
\label{eq:spinorj}
\Gamma^a = [j(\Gamma)]^a = (\tilde j)^{-1} j^a_b \Gamma^b \tilde j.
\end{equation}
Noting that
\begin{equation}
j_a^b \Gamma^a: (\Gamma^0, \Gamma^1, \Gamma^2, \Gamma^3) \rightarrow (\Gamma^0, -\Gamma^1, -\Gamma^2, \Gamma^3),
\end{equation}
\label{eq:gammatransf}
and setting $\tilde j = i \Gamma ^1 \Gamma^2 = (\tilde j)^{-1} = \tilde j^\dagger$ one finds
$(\tilde j)^{-1} \Gamma^a \tilde j = j_b^a \Gamma^b = (j)^{-1}{}_b^a \Gamma^b$ so that \eqref{eq:spinorj} is satisfied as desired.

Since we will use Kaluza-Klein reduction on the $S^1$ to express the four-dimensional fields in terms of three-dimensional fields on a BTZ orbifold, it is useful to express $\tilde j$ in terms of the three-dimensional gamma matrices $\gamma^a$ for $a=(0,1,2)$.  Since each three-dimensional spinor representation is invariant under the action of any $\Gamma^a$, we can choose a basis for the four-dimensional representation in which each $\Gamma^a$ takes a block diagonal form $\begin{pmatrix} \gamma_A^a&0\\ 0&\gamma_B^a\end{pmatrix}$, where the subscripts denote the two spinor representations labelled $A$ and $B$ in section \ref{sec:fermi3}.  In either representation we thus find
\begin{equation}
\tilde j = i \gamma^1 \gamma^2.
\label{eq:pispin}
\end{equation}

The fact that the action of $\hat J_3$ on spinors is ambiguous up to an overall sign implies that there are two natural notions of periodic spinors on the quotient space: those defined by spinors on ${\rm rBTZ} \times S^1$ that are invariant under $\hat J_3$ and those defined by spinors invariant under $-\hat J_3$.  As a result, any use of the terms periodic and anti-periodic spinors is generally a choice of convention as these terms can become well-defined only after making an arbitrary choice of this sign\footnote{This case is like what one often sees in flat space, where the discrete isometry is a special case of a continuous isometry. Continuous isometries lift uniquely from $O(d-1,1)$ to the associated spin group and can be used to define natural notions of periodic and anti-periodic spinors.}.  Our convention in this work will be to use $\hat J_3$ as defined by \eqref{eq:pispin}. In contrast, consider the $\phi$-translation used to construct ${\rm rBTZ}$ as a quotient of AdS$_3$.  This latter isometry already preserves the vielbein \eqref{eq:eamu}, so  we may take the corresponding extra action $j$ on spinor indices to be trivial.

\subsection{Computing $\expval{T_{kk}}$}
\label{subsec:Tkk}
For bosonic fields one defines the source for the Einstein equations by
\begin{equation}
\label{eq:bTmn}
T_{\mu \nu} = \frac{-2}{\sqrt{g}} \frac{\delta S_\textrm{matter}}{\delta g^{\mu \nu}}.
\end{equation}
For spinor fields, as in \eqref{eq:dJAG}, the Lagrangian is best written in terms of the vielbein so that this source is defined via
\begin{equation}
\label{eq:spinT}
T_{\mu \nu} = -\frac{1}{\det(e)} \frac{\delta S_{\textrm{matter}}}{\delta e^\lambda_a}\left(\delta^\lambda_\mu e_{a\nu} + \delta^\lambda_\nu e_{a\mu} \right).
\end{equation}
Note that \eqref{eq:spinT} reduces to \eqref{eq:bTmn} when the vielbein appears in the action only through $g^{\mu \nu} = e^\mu_a \eta^{ab} e^\nu_b$.  Evaluating \eqref{eq:spinT} for a spinor field yields the Belinfante stress-energy tensor
\cite{freedman2012supergravity, DiFrancesco:1997nk}
\begin{equation}
\label{eq:BTmn}
T_{\mu \nu} = \frac{i}{2} \biggl[  \overline{\psi} ( \gamma_{(\mu} D_{\nu)} \psi) + (\overline{D}_{(\nu} \overline{\psi})  \gamma_{\mu)} \psi \biggr],
\end{equation}
where $D_\mu \equiv \partial_\mu + \frac{1}{2} \omega_\mu^{ab} \Sigma_{ab}$ is the covariant derivative with spin connection $\omega_\mu^{ab}$ and $\overline{D_\nu} \equiv \partial_\mu - \frac{1}{2} \omega_\mu^{ab} \Sigma_{ab}$.  This tensor is both symmetric and conserved.

Having understood the action $\hat J_3$ on spinors of the appropriate  ${\rm rBTZ}\times S^1$ isometry $J_3$ in section \ref{subsec:viel} above, we can use the method of images to compute the expectation value $\expval{T_{kk}}$ much as for scalars.
The critical relation is
\begin{equation}
\label{eq:spinimageKKZBO}
\psi (x) \equiv \frac{1}{\sqrt{2}} \left(\tilde  \psi(x) + \left[\hat J_3 \tilde \psi \right]( x ) \right),
\end{equation}
where we have chosen signs such that \eqref{eq:spinimageKKZBO} is a periodic spinor under our $\hat J_3$.  Recall also that   $\left[\hat J_3 \tilde \psi \right]( x ) = j \tilde \psi (J_3 x)$.
Inserting this into \eqref{eq:BTmn} and taking expectation values in the ${\rm rBTZ}\times S^1$ Hartle-Hawking state again yields 4 terms.  Two of these are the expectation values of $\expval{ T_{kk}}$ on ${\rm rBTZ}\times S^1$ at $x$ and at $J_3x$, which for $x$ on the horizon must again vanish by symmetry in the Hartle-Hawking state. We thus need only compute the remaining cross-terms associated with distinct points $x, J_3x$ in the covering space. As for scalars, in the full four-dimensional stress-energy tensor such terms are manifestly non-singular for all $x$.

However, again as for scalars it will be useful to Kaluza-Klein reduce our spinor to a tower of spinors on AdS$_3$ by decomposing $\psi$ into Fourier modes $e^{ip\theta}$ on the internal $S^1$ (we assume periodicity on this circle).  The resulting three-dimensional spinors have masses that are again given by \eqref{eq:KKmass}. As noted earlier, this reduction yields three-dimensional spinors in both representations.

For each $p$ we may write the associated 3D stress-energy tensor $\langle T_{kk}\rangle_p$ on the horizon in terms of the 3D spinor propagator $S_{BTZ}{}^\alpha_{\beta'} (x,x') = \langle  \psi^\alpha(x) \bar \psi_{\beta'}(x') \rangle_{BTZ}$ for a fermion of the appropriate effective mass and choice of spinor representation in the BTZ Hartle-Hawking state
\begin{equation}
\label{eq:Tp2}
\expval{T_{\mu \nu}}_p = (-1)^p \frac{i}{2} \lim_{x \rightarrow x'} \sum_{A,B}  \textrm{Tr} \left[ \gamma_{( \mu} D_{\nu )} S_{BTZ} (x, J_3 x') \tilde j + \tilde j \overline{D_{(\nu' }} S_{BTZ} (J_3 x, x')  \gamma_{\mu )} \right],
\end{equation}
where $D_{\nu'}$ denotes a covariant derivative on the second argument that acts on the associated spinor indices and $\Sigma_{A,B}$ indicates that the right-hand-side adds togeher the contributions from the two three-dimensional representations. The null-null component of this stress-energy tensor can be rewritten
\begin{equation}
\label{eq:Tp3}
\expval{T_{kk}}_p = (-1)^{p+1} \sum_{A,B}  \Im \{ \Tr \left[ \tilde j \slashed{k} D_k S_{BTZ} (x, J_3 x) \right] \}.
\end{equation}
The propagator $S_{\textrm{BTZ}}$ is further given by an image sum $S_{\textrm{BTZ}} = \sum_{n\in {\mathbb Z}} S_{\textrm{AdS}_3}$ over propagators in AdS$_3$.

Using the maximal symmetry of this spacetime, the AdS$_3$ propagators can be written
\begin{equation}
\label{eq:AdSprop}
S_{{\textrm{AdS}}_3}(x,x') = [\alpha(s) + \slashed{n}\beta(s)] \Lambda(x,x')]
\end{equation}
in terms of the spinor parallel propagator $\Lambda$ along the geodesic connecting $x$ and $x'$,  and two functions $\alpha(s), \beta(s)$ of the associated geodesic distance $s$, and  the tangent vector $n^\mu = \frac{\partial}{\partial x^\mu} s(x,x')$ to this geodesic at $x$.  This construction and the relevant formulae are reviewed in appendix \ref{app:paraprop}, following the same procedure as used by \cite{Muck:1999mh} in the Euclidean case.  In particular, the explicit form of $\alpha(s)$ is given by combining \eqref{eq:gammaalpha}, \eqref{eq:gammalambda}, and \eqref{eq:lambda}, whence $\beta(s)$ then follows from the second line of \eqref{eq:betaalpha}.

However, it remains to find an expression for the spinor parallel propagator $\Lambda$.  This of course depends on our choice of $SO(2,1)$ gauge, and thus on our choice of vielbein.  Rather than compute the result directly for the choice \eqref{eq:eamu} and the relevant geodesics, it is simpler to proceed by noting that, for each $n$ and $x$, the contribution to \eqref{eq:Tp2} must be invariant under the combined action of AdS$_3$ isometries, diffeomorphisms, and internal $SO(2,1)$ gauge transformations and using such transformations to separately simplify the computations for each $n,x$.    In particular, we may fix an auxiliary (say, nonrotating) BTZ coordinate system on AdS$_3$ and then use AdS$_3$ isometries to map the geodesic segment running from $x$ to the relevant image point $x'$ onto the bifurcation surface of the (auxiliary) BTZ horizon.  Using the (non-rotating version of the) vielbein \eqref{eq:eamu}, one then finds the spinor parallel propagator $\Lambda$ to be trivial along such geodesics, with $\Lambda (s) = \mathds{1}$ for all $s$.

To proceed further, it is useful to note that the quotient $\tilde M = ({\rm rBTZ} \times S^1)/J_3$ can also be generated by taking the quotient of AdS$_3$ under the group generated by both an rBTZ $\phi$-translation  (for which AdS$_3/{\mathbb Z} = {\rm rBTZ}$) and a $\pi$ rotation in AdS$_3$ global coordinates.  In terms of the standard embedding coordinates reviewed in appendix \ref{app:geo}, this is a rotation in the $(X_1,X_2)$ plane.  For each $n$, we may choosing the lift of $x$ to AdS$_3$ so that the geodesic from $x$ to the relevant image point $x'$ intersects the axis of this rotation at the midpoint of the geodesic, and we may then choose the isometry moving the geodesic to the birfucation surface of our auxiliary BTZ coordinates to be just a rotation around the same axis followed by a translation along it.  In particular, we may choose this AdS$_3$ isometry to preserve the relevant axis so that the action $\tilde j$ on spinor indices of our extended isometry $\hat J_3$ is unchanged. Noting that \eqref{eq:intj} holds in our auxiliary non-rotating BTZ $SO(2,1)$ frame as well as in the physical one, the expression \eqref{eq:pispin} for $\tilde j$ must continue to hold in this frame as well.  Computing each contribution to \eqref{eq:Tp3} in the associated auxiliary frame and summing over $n$ and the choice of representations then yields
\begin{equation}
\expval{T_{kk}}_p (x) = -4 (-1)^p  \sum_{n \in {\mathds Z}} (\varphi_n)_{\mu} k^\mu  \left( n_\mu k^\mu \right)  \left[ \left( \frac{d}{ds_n} - \frac{1}{2\ell} \coth \frac{s_n}{2\ell} \right) \beta(s_n) \right],
\label{eq:Tkk}
\end{equation}
where $(\varphi_n)_\mu$ is the unit-normalized vector at $x$ whose lift to AdS$_3$ points along the infinitesimal generator of rotations of $X^1$ into $X^2$ when the geodesic from $x$ to its image point $x'$ is lifted to AdS$_3$ so as to intersect the associated rotation axis at its midpoint, and  where $s_n$ is the geodesic distance in AdS$_3$ between $x$ and $x'$.  In particular, both $s_n$ and $\varphi_n$ depend on $n$.

As reviewed in appendix \ref{app:geo}, in terms of our rBTZ coordinates the explicit form of the geodesic distance between two points $x,x'$ in AdS$_3$ can be written
\begin{equation}
\begin{split}
s(x,x') = \ell \cosh^{-1} \biggl( \frac{1}{(UV+1)(U'V'+1)} \left[ (UV-1)(U'V'-1) \, \cosh \left( \frac{r_+ (\phi - \phi')}{\ell}\right) + \right. \\ 2 (UV' + VU') \, \cosh \left( \frac{r_- (\phi - \phi')}{\ell} \right)
\left. + 2 (VU' - UV') \, \sinh \left( \frac{r_- (\phi - \phi')}{\ell}\right) \vphantom{l} \right] \biggr).
\end{split}
\end{equation}
Note that choosing $x$ on the horizon and $x'$ as above imposes $U' = V = 0$, $V' = U$, and $\phi' = -\phi$, simplifying the result to
\begin{align}
s(U,\phi) &= \ell \cosh^{-1} \biggl(  \cosh \left( \frac{2 r_+ \phi }{\ell}\right) +  2 U^2 \, \exp \left( -\frac{2 r_- \phi) }{\ell} \right) \biggr) \nonumber \\
&\equiv \ell \cosh^{-1} \left( 1 + \frac{2 \rho^2}{\ell^2}\right)
\end{align}
where $\rho = \ell \sqrt{\sinh^2 \left( \frac{r_+ \phi}{\ell}\right) + U^2 \exp \left( -\frac{2 r_- \phi}{\ell} \right)}$ is the radial coordinate defined by either $x$ or $x'$ in a global AdS$_3$ coordinate system whose rotation axis orthogonally intersects the midpoint of the geodesic from $x$ to $x'$; in other words, $2\pi \rho$ is the circumference of the circle defined by rotating either $x$ or $x'$ about this axis.

\subsection{General Features}
\label{subsec:lim}

Since the general form of $\expval{T_{kk}}$ is quite complicated, we will compute the details of stress-energy tensor profiles and the associated back-reaction on the metric numerically in section \ref{sec:numerics} below.  However, it is useful to first discuss certain general features of \eqref{eq:Tkk}.

In the scalar case, \cite{Fu:2018oaq} found the expressions to simplify greatly in the limit $r_+(\phi - \phi')/\ell, r_-(\phi - \phi')\ell \rightarrow 0$, which in particular holds for the $n=0$ term near $\phi=0$.  This is even more true in our case, as $\varphi_\mu k^\mu$ vanishes in this limit. For $n=0$ we find
\begin{equation}
\begin{aligned}
\expval{T_{kk}}_{n=0,p} = \frac{U  e^{-3 r_- \phi} \sinh \left(r_+ \phi \right) \left( \rho + \sqrt{1+\rho^2}\right)^{-2m_3(p)}}{8 \pi \rho^5 \left( 1 + \rho^2\right)^2} &\left[ 3 + \left( 6 +4m^2_3(p) \right) \rho^2 + \left( 3 + 4 m^2_3(p)\right) \rho^4  \right.
\\ & \left.  + 6 m_3(p) \rho \sqrt{1 + \rho^2} + 8 m_3(p) \rho^3 \sqrt{1+ \rho^2} \right]
\label{eq:Tkkspinor}
\end{aligned}
\end{equation}
with $\rho$ defined as before and where we have set $\ell$ to $1$.

The expression \eqref{eq:Tkkspinor} is singular at $\rho=0$, or $U=\phi=0$. It is useful to understand this singularity since, as described above, the four-dimensional stress-energy tensor can have at most an integrable singularity at this point.  For $p \neq 0$ and $r_+\neq 0$ the terms $\expval{T_{kk}}_{n=0,p}$ are finite, so any non-integrable singularity must cancel when \eqref{eq:Tkkspinor} is summed over $p$.  This is precisely what occurs in the scalar case studied in \cite{Fu:2018oaq}.

However, in our case the singularity in \eqref{eq:Tkkspinor} is separately integrable for each $p$. Writing $e^{-\frac{r_-\phi}{\ell}}U = \rho \cos \theta$ and $\sinh \frac{r_+\phi}{\ell} = \rho \sin \theta$ yields
\begin{equation}
\expval{T_{kk}}_{n=0,p} = \frac{3 \sin \theta \cos \theta}{8 \pi \rho^3} \left(1  - 2\frac{r_-}{r_+} \cos \theta \rho + O(\rho^2)\right)
\label{eq:Tkkspinorexp}
\end{equation}
at small $\rho$.  The result simply vanishes for any $U$ at $\phi =0$ ($\sin \theta=0$) for any $U$, or for any $\phi$ at $U=0$ ($\cos \theta=0$). Integrating over $U$ at fixed $\phi$ thus raises no issues.  Integrating over both $\phi$ and $U$ also yields a finite result since $\int d \theta \sin \theta \cos \theta =0 = \int d \theta \sin \theta \cos \theta $, so that integral of the explicit terms in \eqref{eq:Tkkspinorexp} also vanish, and all other terms give finite results due to the fact that the measure $dUd\phi \propto \rho d\rho d\phi (1 + O(\rho^2))$ supplies an additional factor of $\rho$.

The expression \eqref{eq:Tkkspinor} has many similarities to the scalar expression \eqref{eq:Tkkboson}. One might expect a particularly simple relation between the two in the large mass limit where occupation numbers are small and quantum effects are suppressed so that the choice of bosonic vs. fermionic statistics is unimportant. But the kinematic structure of the expressions remains different in that limit, associated with the non-trivial action $\tilde j$ of the isometry $\hat J_3$ on fermionic indices. In particular, at large $m_3$ the spinor result \eqref{eq:Tkkspinor} yields
\begin{equation}
\expval{T_{kk}}_{n=0,\psi} = \frac{U \sinh \left( r_+ \phi \right) e^{-3 r_- \phi}\left( \rho + \sqrt{1 + \rho^2}\right)^{-2m_3}}{2 \pi \rho^3 \left( 1 + \rho^2\right)} m_3^2 + \mathcal{O} (m_3),
\end{equation}
while $\Delta \rightarrow 1 + m$ in the bosonic expressions \eqref{eq:Tkkboson} gives
\begin{equation}
\expval{T_{kk}}_{n=0,\phi} = \frac{U^2 e^{-4 r_- \phi}\left( \rho + \sqrt{1 + \rho^2}\right)^{-2m}}{8 \pi \rho^3  \left( 1 + \rho^2\right)^{3/2}}m^2 + \mathcal{O} (m).
\end{equation}
The fact that the denominators differ by a factor of $4$ may be ascribed to the fact that we consider a single real scalar and a four-component spinor (from the four-dimensional point of view).  However, the other discrepancies reflect the difference in kinematics.  We find similar differences in the limit of large $U$ with $m$ fixed, which yields
\begin{equation}
\frac{\expval{T_{kk}}_{n=0,\psi}}{\expval{T_{kk}}_{n=0,\phi}} = 4 \sinh \left( r_+ \phi\right)  + \mathcal{O} (1/U^2).
\end{equation}
It is interesting that the above expressions for our fermion field all change signs under $\phi \rightarrow -\phi$.  This stands in marked contrast to the scalar results whose signs are generally $\phi$-independent\footnote{Though with specially engineered boundary conditions \cite{Fu:2018oaq} found cases where the sign of the integrated scalar stress-energy depends on $\phi$ as well.}. This odd-parity behavior turns out to arise from the non-rotating limit $r_-(\phi - \phi')\ell \rightarrow 0$ regardless of whether a similar limit is taken for $r_+$.  Indeed, for non-rotating BTZ there is no preferred sign of the unit-vector $\varphi_\mu$ and the symmetry of non-rotating BTZ under $\phi \rightarrow -\phi$ requires $(\varphi_n)_\mu k^\mu \rightarrow - (\varphi_n)_\mu k^\mu $.  Since in that case all other factors in $\langle T_{kk}\rangle$ are even, the integrated stress-energy becomes an odd function of $\phi$ and the average over the full horizon must vanish.  But this symmetry is broken by rotation and, indeed, we will find below that for non-zero angular velocity the average of $\expval{T_{kk}}$ over the full horizon is non-zero.

\subsection{Numerical Results}
\label{sec:numerics}
It remains to study $\expval{T_{kk}}$ and $\int dU \expval{T_{kk}}$ in detail. For this task we resort to numerics and follow the same basic strategy as in \cite{Fu:2018oaq}. We will phrase our results in terms of the dimensionless quantity $\ell \expval{T_{kk}} \propto \Delta V$. As above, our results are for the effective three-dimensional Kaluza-Klein-reduced stress-energy tensor. We impose a cutoff $N$ on the number of Kaluza-Klein modes over which we sum and also regulate the functions near the (integrable) singularity at $U=0, \phi=0$.  We choose our cutoffs such that our answers do not change significantly when these cutoffs are altered.

In particular, our numerical expressions are computed via
\begin{equation}
\expval{T_{kk}}_{numerical} = \sum_{p = -N}^{N} (-1)^p f (U, m (p)) + (-1)^{N+1}f (U, m( N + 1))
\end{equation}
where we have added an extra term so as to sum over an even number of terms, $N$ of which have an additional sign change. In computing $\left(\int dU \expval{T_{kk}} \right)_{\textrm{numerical}}$ we integrate the stress-energy tensor only over $\abs{U} > \epsilon$. We will interpolate between the origin and $U = \epsilon$ with a linear approximation for the spinor and a constant for the scalar, as shown in Figure \ref{fig:approx} and integrate the interpolating function for $\abs{U} < \epsilon$. We choose our spinors to be periodic under $\hat J_3$ such that we get a positive overall contribution to the stress-energy tensor.

\begin{figure}[t]
\begin{center}
\includegraphics[width=0.7\textwidth]{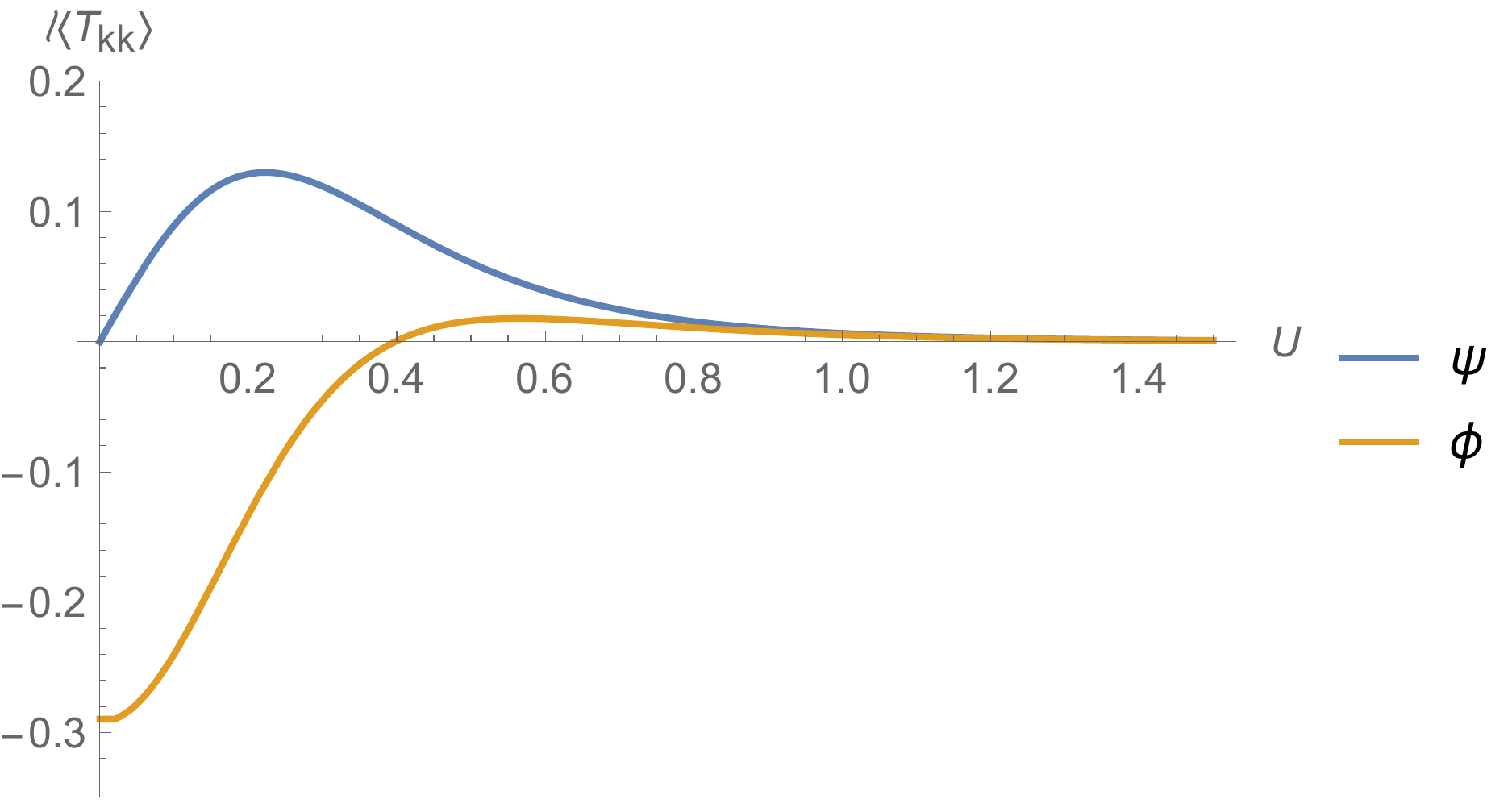}
\end{center}
\caption{Plot of $\ell \expval{T_{kk}}$ vs. $U$ with $k^\mu \partial_\mu =\partial_U$ for a spinor field $\psi$ and a scalar field $\phi$. Both particles have mass $m = 1$ in units of inverse $\ell_{AdS}$. We have chosen $r_+ = 1$ and $r_- = 1/2$ in the same units, as well as $\phi = 0.1$. Here the cutoff $\epsilon = 0.02$, the sum over BTZ images has been done to $n = 3$, and the sum over KK modes has been performed to $N = 50$, with $\ell/R_{S^1} = \sqrt{10}$.}
\label{fig:Tkk}
\end{figure}

\begin{figure}[t]
\begin{center}
\includegraphics[width=0.4\textwidth]{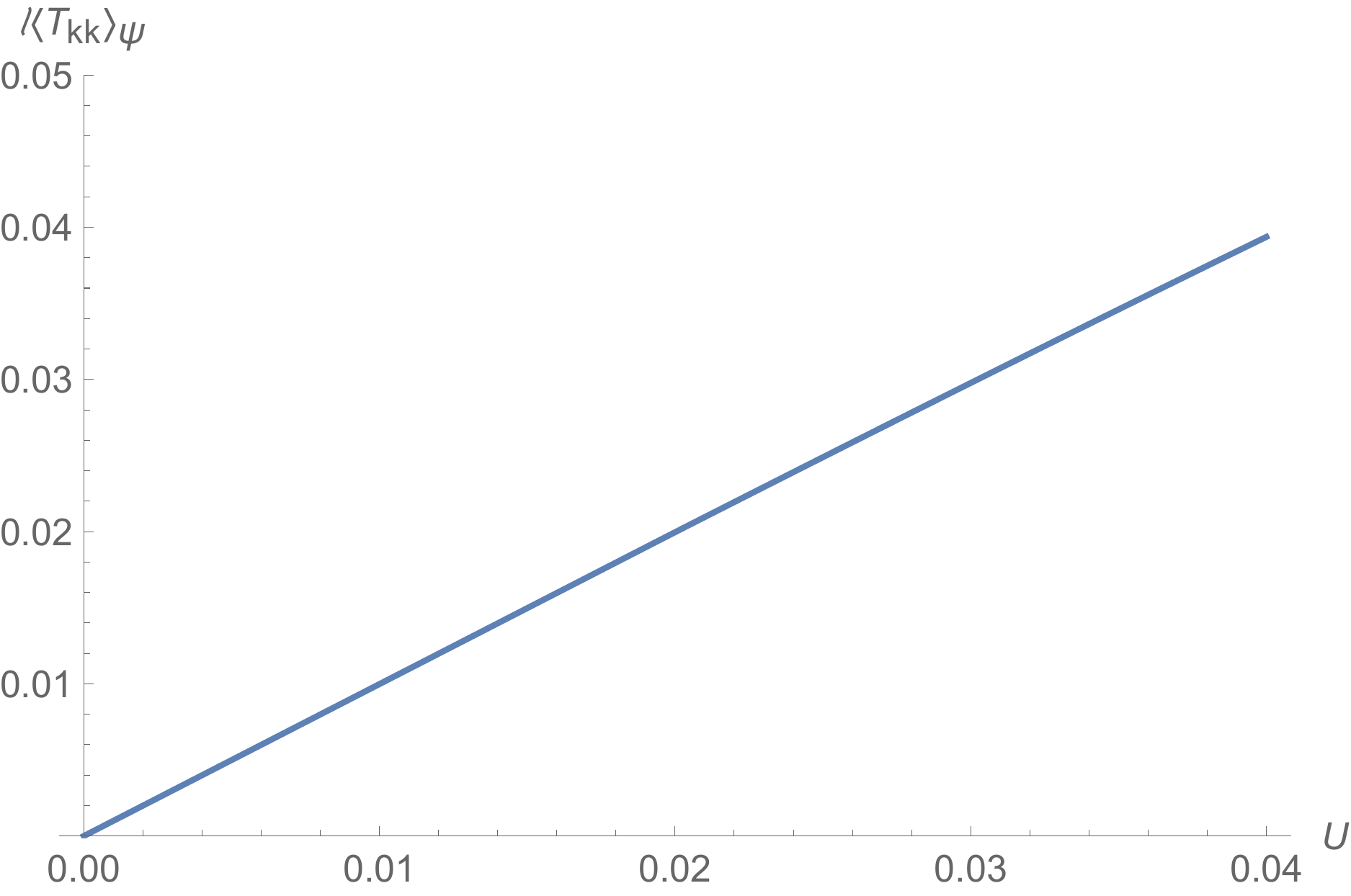}
\includegraphics[width=0.4\textwidth]{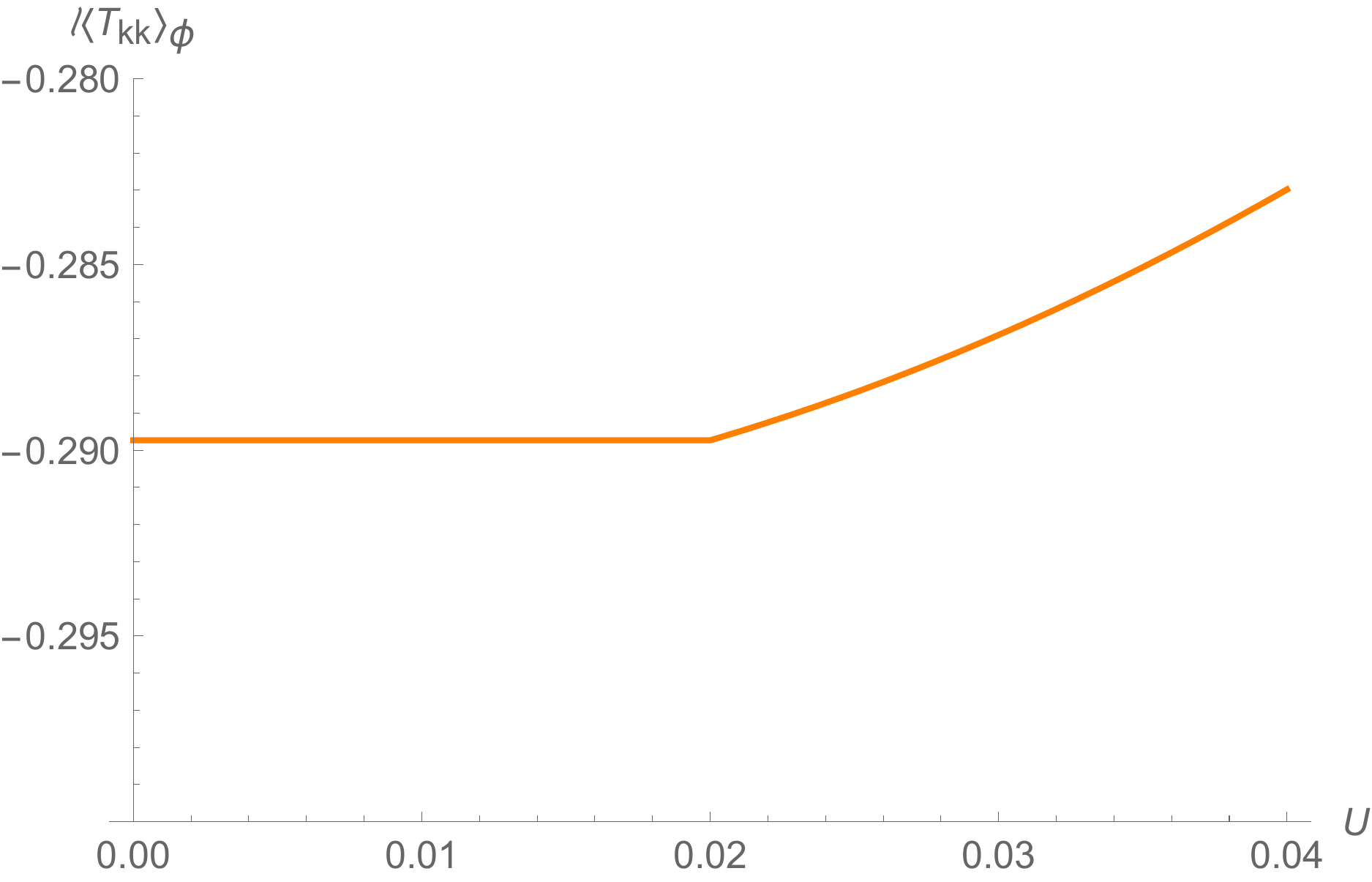}
\end{center}
\caption{Details of the interpolation between the origin and $U = \epsilon = 0.02$ for the spinor and scalar in Figure \ref{fig:Tkk}.}
\label{fig:approx}
\end{figure}

Some results for the dimensionless quantity $\ell \expval{T_{kk}}$ are shown in figures \ref{fig:Tkk} and \ref{fig:approx}. Figure \ref{fig:Tkk} shows the contributions to the stress-energy tensor for a spinor and scalar of the same mass at a particular value of $\phi$. Notably, the spinor contribution is everywhere positive, as opposed to the scalar contribution which changes sign. Figure \ref{fig:approx} shows the details of the interpolation of $\expval{T_{kk}}$ for small $U$. As the spinor stress-energy tensor vanishes at $U=0$, a linear interpolation from the origin was used, while a constant interpolation was used for the scalar case.

\begin{figure}[t]
\begin{center}
\includegraphics[width=0.8\textwidth]{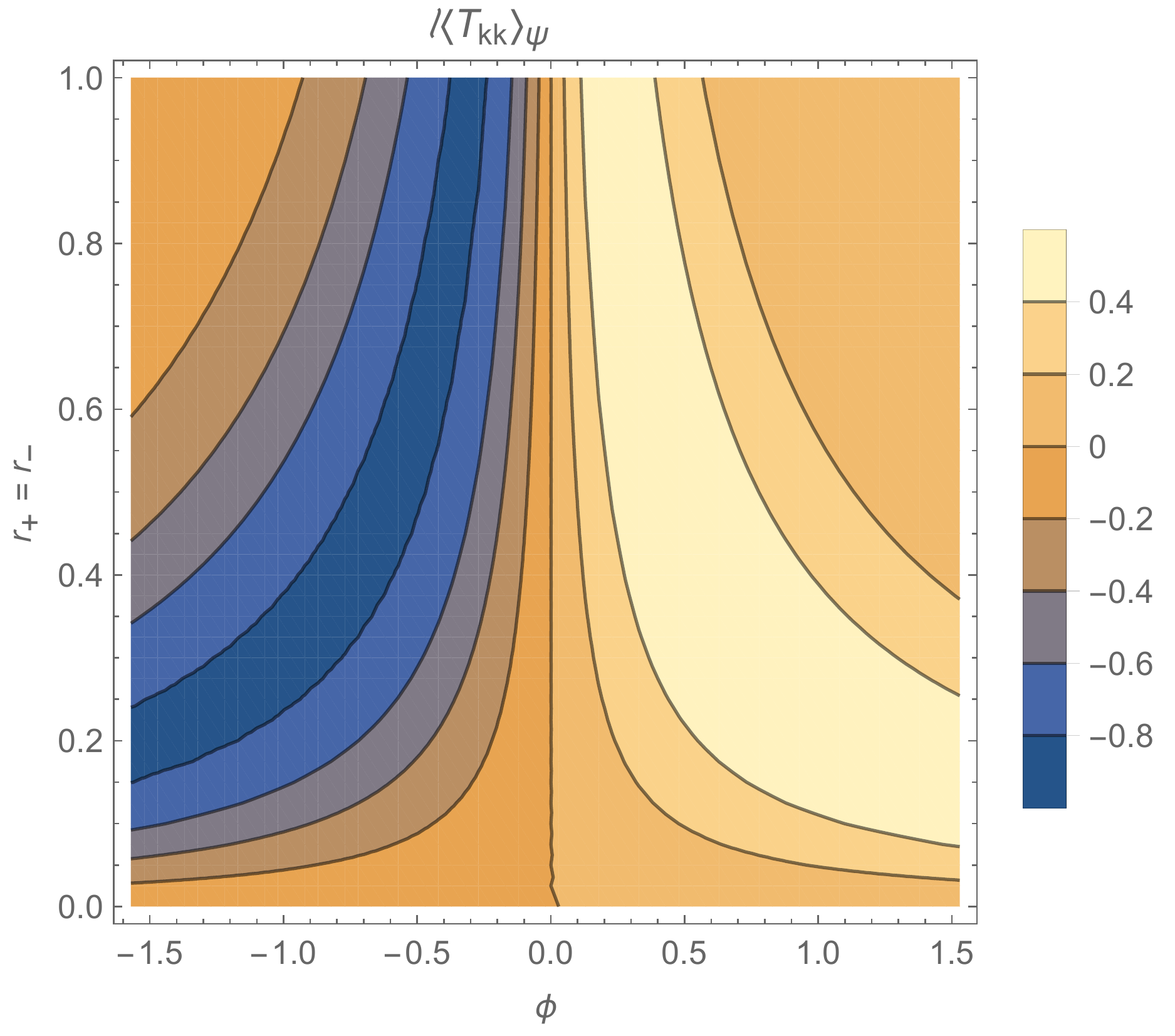}
\end{center}
\caption{Plot of $\ell  \expval{T_{kk}}$ at extremality for a spinor $\psi$ at various values of $r_+ = r_-$ and $\phi$. The four dimensional mass $m = 1$ and $\ell/R_{S^1} = \sqrt{10}$. The cutoff $\epsilon = 0.1$, the sum over Kaluza-Klein modes was performed to $N=20$, and the sum over BTZ images was performed to $n=3$.}
\label{fig:extremalTkk}
\end{figure}

\begin{figure}[t]
\begin{center}
\includegraphics[width=0.7\textwidth]{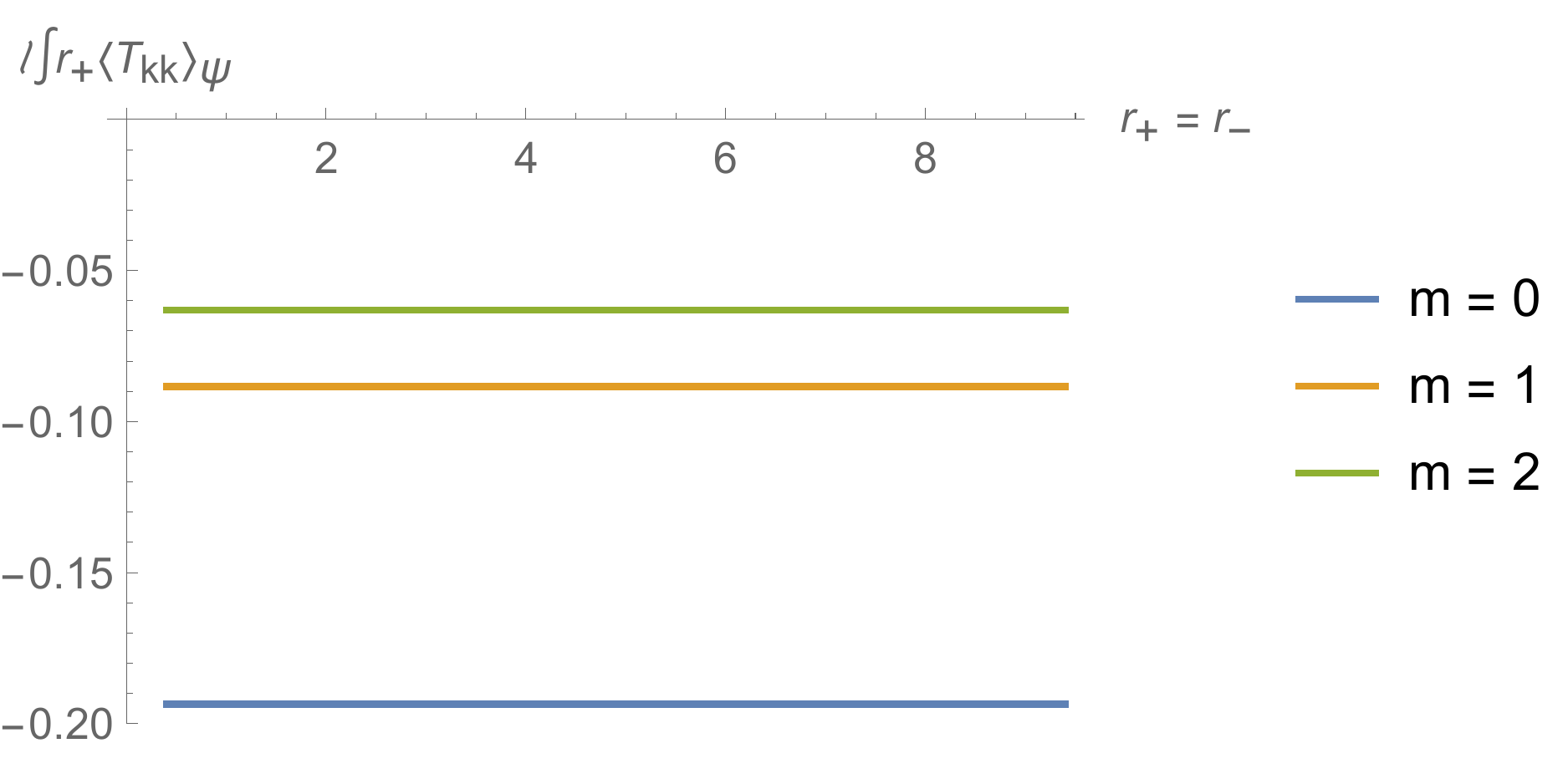}
\end{center}
\caption{Plot of the quantity $ \ell \int_0^\infty \int_{-\pi/2}^{\pi/2} d\phi dU r_+ \expval{T_{kk}}$ at extremality for $m = (0, 1, 2)$ and $\ell/R_{S_1} = 10$. The cutoff $\epsilon = 0.01$, the sum over Kaluza-Klein modes modes was performed to $N=200$, and the sum over BTZ images was performed to $n = 3$. This quantity appears constant for all values of $r_+$, up to corrections at small $r_+$ for the small value of $N$. Relevant values of the four-dimensional mass $m$ are indicated above.}
\label{fig:zeroMode}
\end{figure}

\begin{figure}[t]
\begin{center}
\includegraphics[width=0.7\textwidth]{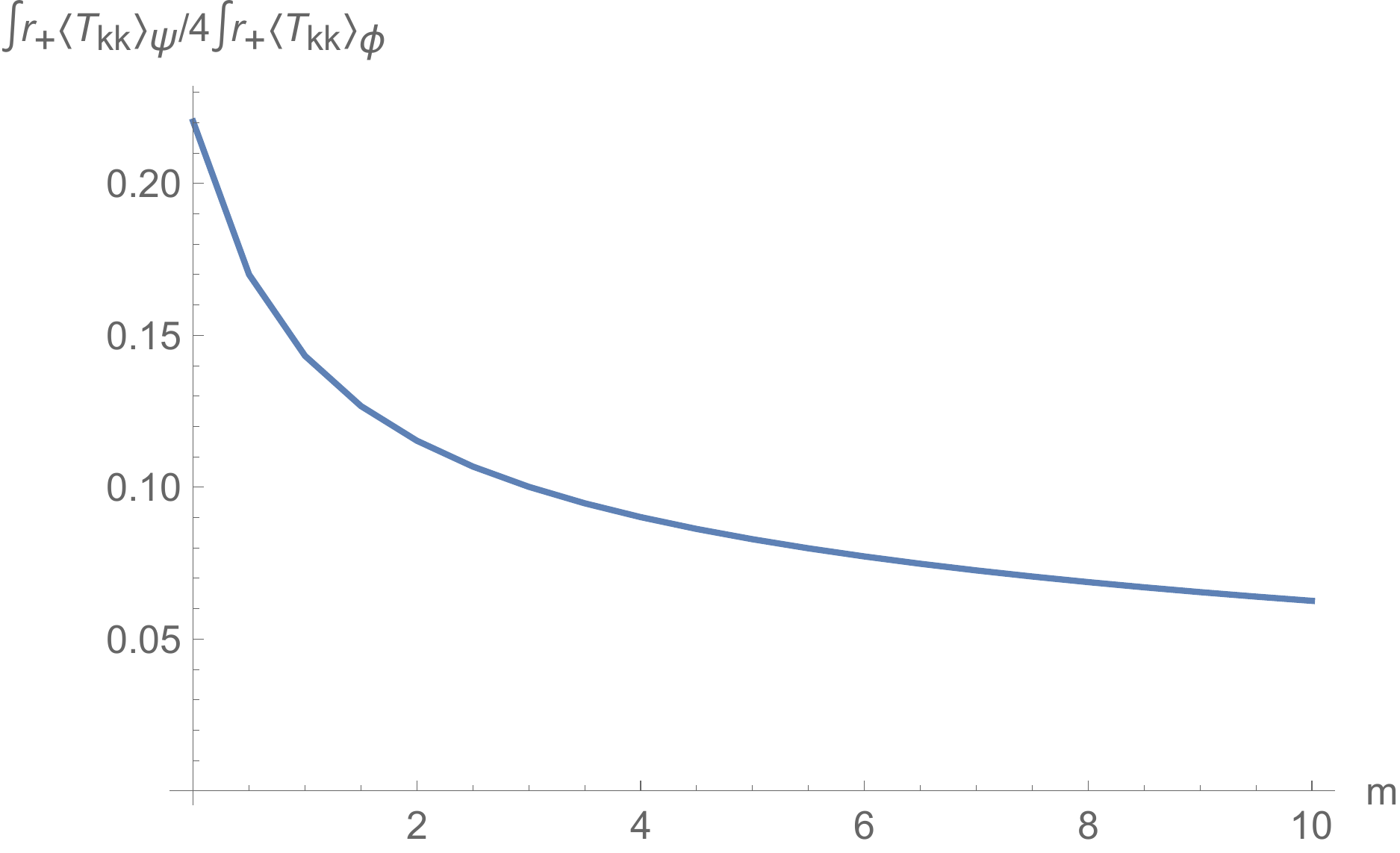}
\end{center}
\caption{Plot of the ratio of the quantity $\int_0^\infty \int_{-\pi/2}^{\pi/2} d\phi dU r_+ \expval{T_{kk}}$ at extremality for one four-dimensional complex spinor field and four four-dimensional real scalar fields of the same four-dimensional mass $m$. $\ell/R_{S_1} = 10$ with a cutoff $\epsilon = 0.01$. The sum over Kaluza-Klein modes was performed to $N=1000$ and the sum over BTZ images was performed to $n = 5$. The spinor contribution to the integrated stress-energy tensor is always significantly less than that of the scalars with an equivalent number of degrees of freedom.}
\label{fig:zmComp}
\end{figure}

We also numerically calculate the  contributions to $\int dU \expval{T_{kk}}$ at extremality in figure \ref{fig:extremalTkk} for various values of $\phi$ and $r_+ = r_-$. As opposed to the scalar contribution, the spinor contribution generically changes sign at different $\phi$ for a given $r_+$. In the scalar case, this phenomenon must be engineered \cite{Fu:2018oaq} by requiring some KK modes to have the $(-)$ boundary condition in \eqref{eq:cft}. Figure \ref{fig:zeroMode} plots the integral of $\langle T_{kk} \rangle$ over both $U$ and $\phi$  at extremality as a function of the radius $r_+ = r_-$.  The physical importance of this quantity is that, as discussed in section \ref{subsec:KKZBO}, it is proportional to $T \expval{V}_{\textrm{average}}$ (see \eqref{eq:DVav}).  As in \cite{Fu:2018oaq}, we find numerically that this function is independent of $r_+$.  We thus find a large average time-advance
$\expval{V}_{\textrm{average}} \propto 1/T$ as $T \rightarrow 0$ in the extreme limit $r_- \rightarrow r_+$, suggesting that a non-perturbative treatment may lead to an eternally traversable wormhole as in \cite{Fu:2018oaq}, at least in the presence of some large parameter that controls quantum fluctuations relative to the mean. Figure \ref{fig:zmComp} compares the relative size of this quantity at extremality for one four-dimensional complex scalar field and four four dimensional real scalar fields, which as mentioned before, have the same number of degrees of freedom. For all values of the three-dimensional mass $m$, the spinor contribution to the stress-energy tensor is less than that of the equivalent scalar fields, owing to cancellations from the spinor kinematic structure.

\section{Conclusion}
\label{sec:conclude}

We have studied the contributions of bulk spinors to the integrated null stress-energy tensor of a $\mathbb{Z}_2$ quotient of rBTZ $\times \; S^1$ known as the Kaluza-Klein zero-brane orbifold (KKZBO).  As in the scalar case, the $\mathbb{Z}_2$ quotient is associated with a sign that controls the periodicity of the bulk field as well as the overall sign in the null stress-energy $\langle T_{kk}\rangle$ on the horizon.

The fact that spinor fields carry Lorentz indices leads to notable differences from the scalar case. The spinor $\expval{T_{kk}}$ includes an overall factor of $\varphi_\mu k^\mu$, where $\varphi$ is a vector field defined by the $\mathbb{Z}_2$ quotient operation.  This $\varphi_\mu k^\mu$ is an odd function of $\phi$ for non-rotating BTZ, so in that case the spinor $\langle T_{kk} \rangle$  is odd as well and integrates to zero over the $\phi$-circle.  As a result, at least in the limit of large black hole radius where the Green's function \eqref{eq:HGF} relating $\langle T_{kk} \rangle$ to the generator-dependent time delay $\langle \Delta V\rangle$ becomes short-ranged, without rotation one finds the wormhole to be traversable when entered across one half of the $\phi$-circle (say for $\phi \in (0,\pi)$) but to remain non-traversable when entered across the other half-circle.

This overall factor of $\varphi_\mu k^\mu$ is associated with the fact that the simplicity of the model in four dimensions means that both possible three-dimensional representations of the Clifford algebra enter with equal weight.  Terms that for each representation are not proportional to $\varphi_\mu k^\mu$ have opposite signs in the two representations and cancel. As a result, a more chiral construction that caused the two representations to enter on a less equal footing would not have this property.

While the above $\phi \rightarrow -\phi$ (anti-)symmetry is broken at non-zero angular velocity, the sign of $\expval{T_{kk}}$ still varies with $\phi$ and the average over the $\phi$-circle is correspondingly reduced relative to the scalar case.  But choosing the aforementioned sign correctly still makes the average negative (and thus also $\expval{V}_{\textrm{average}}$). As in the scalar case \cite{Fu:2018oaq}, the quantity $T \expval{V}_\textrm{average} <0$ is nonzero at $T=0$, so that the time delay $-\expval{V}_\textrm{average} \propto 1/T$ becomes large. At least in the presence of some large parameter that controls quantum fluctuations relative to the mean, this suggests that a non-perturbative treatment may lead to an eternally traversable wormhole as in \cite{Maldacena:2018gjk}. It also interesting that, again as in the scalar case studied in \cite{Fu:2018oaq}, numerical results suggest $T \expval{V}_\textrm{average}$ to be independent of $r_+ = r_-$ at extremality, though with values noticeably smaller than in the scalar case (see figure \ref{fig:zmComp}) due to partial cancellations  associated with the variation in sign with $\phi$ described above.

Such results in particular make clear that, even though extreme BTZ preserves certain supersymmetries, models with $N=1$ bulk supersymmetry do not lead to special cancellations in $\expval{T_{kk}}$ between bosons and fermions on the extreme KKZBO spacetime. This lack of cancellations is especially manifest at large mass $m$ where AdS bulk supersymmetry relates fermions and bosons of nearly equal masses. It should not be a surprise since the non-vanishing of  $\expval{T_{kk} }_{\textrm{KKZBO}}$ is manifestly due to the breaking of the BTZ Killing symmetry by the $\mathbb{Z}_2$ quotient used to build KKZBO from extreme BTZ.  Since this Killing symmetry is part of the supersymmetry algebra on extreme BTZ, the quotient must break supersymmetry at the same level.

A natural next step for further investigation would be to study fields of even higher spin, and in particular to better understand what kinematic factors might arise in such cases. Expressions for the vector propagator for both massive and massless particles on AdS$_3$ were given in \cite{allen1986a} and can be used to compute the stress-energy of such fields on the KKZBO spacetime in direct analogy to the computations performed here.  Further extensions to spin-3/2 and spin-2 fields would then allow one to study gravitino and graviton back-reaction to our wormhole geometry.   It would also be interesting to study the back-reaction of signals passing through our wormholes as was done for GJW wormholes in \cite{Maldacena:2017axo,Caceres2018,Hirano:2019ugo,Freivogel:2019whb} to understand how differences from the scalar case above affect limits on the amount of information that can be transmitted.

\section*{Acknowledgements}
We would like to thank Zicao Fu, Brianna Grado-White, and Aron Wall for useful discussions. This work was supported in part by NSF grant PHY1801805 and in part by funds from the University of California.

\appendix

\section{Geodesic Length in Kruskal-like Coordinates}
\label{app:geo}

We start with AdS$_3$ in Rindler coordinates. The metric in the embedding space $\mathds{R}^{2,2}$ is
\begin{equation}
ds^2 = -dT_1^2 - dT_2^2 + dX_1^2 + dX_2^2.
\end{equation}
The usual Rindler patch has a horizon at $r = r_+$, but to describe a rotating black hole solution we requires two horizons $r_+$ and $r_-$. The embedding functions for this geometry are given by
\begin{align}
T_1 &=  \ell \sqrt{\alpha} \cosh \left( \frac{r_+ \phi}{\ell} - \frac{r_- t}{\ell^2}\right)\nonumber \\
X_1 &=  \ell \sqrt{\alpha} \sinh \left( \frac{r_+ \phi}{\ell} - \frac{r_- t}{\ell^2}\right)\nonumber \\
T_2 &= \ell \sqrt{\alpha-1} \sinh \left( \frac{r_+ t}{\ell^2} - \frac{r_- \phi}{\ell} \right) \nonumber \\
X_2 &=  \ell \sqrt{\alpha-1} \cosh \left( \frac{r_+ t}{\ell^2} - \frac{r_- \phi}{\ell} \right)
\end{align}
for $\alpha \equiv \frac{r^2 - r_-^2}{r_+^2 - r_-^2}$. For this problem, we would like a coordinate reparametrization that makes the null directions manifest, so we use the coordinates $(U,V,\phi)$, where $U$ and $V$ are defined to be
\begin{align}
U &= \exp \left[ \kappa (t + r_*) \right] \nonumber \\
V &= \exp \left[ - \kappa (t - r_*) \right]
\end{align}
where $\kappa = \frac{r_+^2 - r_-^2}{\ell ^2 r_+}$ and $r_* = \frac{\ell^2}{2 \kappa} \left( \frac{\sqrt{r_+^2 - r_-^2} - \sqrt{r^2-r_-^2}}{\sqrt{r_+^2 - r_-^2} + \sqrt{r^2-r_-^2}} \right)$. In our new Kruskal-like coordinates, and using the corotating coordinates $\phi \rightarrow \phi - \frac{r_-}{\ell r_+}t$, the AdS$_3$ embedding functions become
\begin{align}
T_1 &=  \ell \left( \frac{U + V}{1 +UV} \cosh \frac{r_- \phi}{\ell} - \frac{U - V}{1 +UV} \sinh \frac{r_- \phi}{\ell} \right)\nonumber \\
X_1 &= \ell \left( \frac{U - V}{1 +UV} \cosh \frac{r_- \phi}{\ell} - \frac{U + V}{1 +UV} \sinh \frac{r_- \phi}{\ell} \right)\nonumber \\
T_2 &= \ell \frac{1-UV}{1+UV} \cosh \frac{r_+ \phi}{\ell} \nonumber \\
X_2 &= \ell \frac{1-UV}{1+UV} \sinh \frac{r_+ \phi}{\ell}.
\label{nullreparam}
\end{align}
The metric that results from these embedding functions is
\begin{equation}
ds^2 = \frac{1}{(1+UV)^2} \left( -4\ell^2 dU dV + 4 \ell r_- (U dV - V dU) + (r_+^2 (1 - UV)^2 + 4 U V r_-^2) d \phi^2 \right).
\end{equation}
Under an identification $\phi \sim \phi + 2 \pi$, this metric becomes the BTZ metric with $r_+^2+r_-^2 = \ell^2M$.

With our embedding functions \eqref{nullreparam}, one can also derive an expression for the geodesic distance $s$ in our AdS$_3$ Kruskal-like coordinates using the geodesic distance in the embedding spacetime. The result is \cite{Nastase:2015wjb}
\begin{align}
s (x, x') &= \ell \cos^{-1} \sigma (x, x') \ \ \ \ \ \ \ \ \ \ \ \ s \textrm{ timelike} \nonumber\\
s (x, x') &= \ell \cosh^{-1} \sigma (x, x')  \ \ \ \ \ \ \ \ \ \ s \textrm{ spacelike}
\end{align}
The geodesic we will be concerned with for our calculations is the spacelike case. This $\sigma (x, x')$ is the same quantity that appears in the calculation of the scalar Green's function and is given by
\begin{equation}
\sigma (x, x') = \frac{1}{\ell^2} \left[ T_1 (x) T_1 (x') + T_2 (x) T_2 (x') - X_1 (x) X_1 (x') + X_2 (x) X_2 (x') \right].
\end{equation}
Substituting our Kruskal-like embedding functions, we arrive at an expression for the geodesic distance in AdS$_3$
\begin{equation}
\label{eq:explicitgeodesic}
\begin{split}
s = \ell \, \cosh^{-1} \biggl( \frac{1}{(UV+1)(U'V'+1)} \left[ (UV-1)(U'V'-1) \, \cosh \left( \frac{r_+ (\phi - \phi')}{\ell}\right) + \right. \\ 2 (UV' + VU') \, \cosh \left( \frac{r_- (\phi - \phi')}{\ell} \right)
\left. + 2 (VU' - UV') \, \sinh \left( \frac{r_- (\phi - \phi')}{\ell}\right) \vphantom{l} \right] \biggr).
\end{split}
\end{equation}
As a consistency check, the norm of this geodesic, $n_\mu \equiv \partial_\mu s$, satisfies $n^2 = 1$, as it should for a spacelike geodesic, $\nabla_\mu n_\nu \equiv n^\mu \nabla_\mu n_\nu = 0$, and the constraints on the bitensor of parallel transport $g_\nu^{\nu'}$ in Appendix C of \cite{allen1986a}.

\section{Spinor Propagator in AdS$_d$}
\label{app:paraprop}

We follow the derivation given in \cite{Muck:1999mh} for the parallel propagator in $\mathds{H}_d$, with the necessary few changes required for analytic continuation to AdS$_d$ pointed out in footnotes. The calculation of $\expval{T_{\mu \nu}}$ for timelike geodesics in $d = 4$ can be found in \cite{ambrus}.

We start with defining the norms of the AdS$_d$ geodesic, which are given by
\begin{align}
n_\mu &= \partial_\mu s (x, x') \nonumber \\
n_{\mu'} &= \partial_{\mu'} s (x,x').
\end{align}
We also define the bitensor of parallel transport $g_\mu^{\nu'}$, which takes vectors between the two tangent spaces defined at $x$ and $x'$. In particular, $n_\mu = -g_\mu^{\nu'} n_{\nu'}$, as the geodesic norm at a point is in the opposite direction of the geodesic length.

We also define two functions $A$ and $C$ relating the derivatives of $n_\mu$ and $n_{\mu'}$ to similarly-indexed quantities\footnote{In the case of timelike geodesics, these differ from \cite{allen1986a} by the transform $g_{\mu \nu} \rightarrow -g_{\mu \nu}$ and $g_{\mu \nu'} \rightarrow -g_{\mu \nu'}$}
\begin{align}
\nabla_\mu n_\nu &= A (g_{\mu \nu} - n_\mu n_\nu ) \nonumber \\
\nabla_\mu n_{\nu'} &= C (g_{\mu \nu'} + n_\mu n_{\nu'}).
\label{eq:AC}
\end{align}
For spacetimes with negative curvature, $A = \frac{1}{\ell}\coth \left( \frac{s}{\ell} \right)$ and $C = -\frac{1}{\ell} \csch \left( \frac{s}{\ell} \right)$\footnote{In \cite{Muck:1999mh}, these functions in $\mathds{H}_d$ are instead given by $A = \frac{1}{\ell}\cot \left( \frac{s}{\ell} \right)$ and $C = -\frac{1}{\ell} \csc \left( \frac{s}{\ell} \right)$}. By inverting the second equation and substituting the first, we find
\begin{equation}
\nabla_\mu g_{\nu \lambda'} = - (A + C) (g_{\mu \nu} n_{\lambda'} + g_{\mu \lambda'} n_{\nu}).
\end{equation}
In order to properly treat spinors at two disconnected points, we also need to write a spinor parallel propagator $\Lambda^{\alpha'}_{\beta}$, which transports a spinor from $x$ to $x'$ as
\begin{equation}
\Psi' (x')^{\alpha'} = \Lambda (x', x)^{\alpha'}_{\beta} \Psi (x)^{\beta}.
\end{equation}
The covariant derivatives of $\Lambda^{\alpha'}_{\beta}$ with respect to primed and unprimed coordinates are fixed by the parallel transport of the gamma matrices and \eqref{eq:AC} to be
\begin{align}
D_\mu \Lambda (x, x') &= \frac{1}{2}(A+C)(\gamma_\mu \gamma^\nu n_\nu - n_\mu) \Lambda (x, x') \nonumber \\
D_{\mu'} \Lambda (x, x') &= -\frac{1}{2}(A+C)\Lambda (x, x')(\gamma_{\mu'} \gamma^{\nu'} n_{\nu'} - n_{\mu'}).
\label{eq:DLambda}
\end{align}
In principle, we could define a vierbein, find the exact form of the gamma matrices and the covariant derivative, and solve a system of PDEs for the spinor parallel propagator. However, as for the scalar case, making use of the maximal symmetry of AdS turns out to greatly simplify the derivation \cite{Muck:1999mh}.

The spinor propagator $S^\alpha_{\beta'} (x,x') = \langle  \psi^\alpha(x) \bar \psi_{\beta'}(x') \rangle$ is a solution of the spacelike Dirac equation
\begin{equation}
\left[ (\slashed{D} - m) S(x,x')\right]^\alpha_{\beta'}  = \frac{\delta (x-x')}{\sqrt{-g}}\delta^\alpha_{\beta'}
\label{eq:Dirac}
\end{equation}
with appropriate short-distance singularities.  In the AdS vacuum it will share the maximal symmetry of the spacetime.  It must therefore be of the form
\begin{equation}
S(x,x') = [\alpha(s) + \slashed{n} \beta(s)]\Lambda (x,x'),
\label{eq:ansatz}
\end{equation}
for $\slashed{n} = n_\mu \gamma^\mu$ and some functions of the geodesic distance $\alpha(s)$ and $\beta(s)$. Inserting \eqref{eq:ansatz} into \eqref{eq:Dirac} yields
\begin{equation}
\left[ \left( \alpha' + \frac{1}{2}(d-1)(A+C)\alpha - m \beta \right) \slashed{n} + \left( \beta' + \frac{1}{2} (d-1)(A-C) \beta - m \alpha \right) \right] \Lambda (x, x') = \frac{\delta (x-x')}{\sqrt{-g}}\delta^\alpha_{\beta'},
\end{equation}
where a prime indicates a derivative with respect to $s$. Taking the trace of the above equation gives a set of two coupled differential equations
\begin{align}
\label{eq:betaalpha}
\beta' + \frac{1}{2} (d-1)(A-C) \beta - m \alpha &= \frac{\delta (x-x')}{\sqrt{-g}}, \nonumber \\
\alpha' + \frac{}{2}(d-1)(A+C)\alpha - m \beta  &= 0,
\end{align}
Substituting the second equation into the first and using identities involving $A$ and $C$ from table 1 of \cite{allen1986a}, one finds a second order equation for $\alpha (s)$:
\begin{equation}
\alpha'' + (d-1)A \alpha' - \left[ m^2 + \frac{1}{2}(d-1)(C^2 + AC) - \frac{(d-1)^2}{4\ell^2} \right] \alpha = m \frac{\delta (s)}{\sqrt{-g}}.
\label{eq:alpha}
\end{equation}
\subsection{Solution in Minkowski Space}
In order to find the proper normalization of the solutions in AdS$_d$, we need to find the short distance behavior of $\alpha (s)$, which is just the solution of \eqref{eq:alpha} in Mink$_d$, or equivalently, in the limit $\ell \rightarrow \infty$. \eqref{eq:alpha} becomes
\begin{equation}
\alpha'' + \frac{d-1}{s} \alpha' - m^2 \alpha = m \delta (s).
\end{equation}
The solution to this equation, properly normalized such that the left and right sides agree in the coincident limit, is
\begin{equation}
\alpha (s) = - \left( \frac{m}{2 \pi} \right)^{d/2} s^{1-d/2} K_{d/2-1} (ms)
\end{equation}
where $K_n (z)$ is the modified Bessel function of the second kind. The series expansion of this solution around $s = 0$ is given by
\begin{equation}
\alpha (s) \approx -\frac{m}{4}\pi^{-d/2}s^{2-d}\Gamma\left(\frac{d}{2}-1\right).
\label{eq:ShortDistance}
\end{equation}
\subsection{Solution in AdS$_d$}
To solve \eqref{eq:alpha} in AdS$_d$, we make the substitutions $z \equiv \cosh^2 \left( \frac{s}{2\ell} \right)$ and
\begin{equation}
\label{eq:gammaalpha}
\gamma (z) \equiv \frac{1}{\sqrt{z}} \alpha(z)
\end{equation} to obtain\footnote{For timelike geodesics, we would instead make the same substitutions as in \cite{Muck:1999mh}, with $z \equiv \cos^2 \left( \frac{s}{2\ell} \right)$. }
\begin{equation}
z(1-z) \gamma''(z) + \left(\frac{d}{2} + 1 - (d + 1) z\right) \gamma'(z) + \left( m^2 \ell^2 - \frac{d^2}{4}\right) \gamma (z) = -m \frac{\delta (s)}{\sqrt{-g}},
\end{equation}
where here primes denote derivatives with respect to $z$. This is a hypergeometric equation in $\gamma (z)$, with differential operator
\begin{equation}
H (a,b,c;z) = z(1-z) \frac{d^2}{dz^2} + \left( c - (a + b - 1)z \right) \frac{d}{dz} - ab,
\end{equation}
with $a = \frac{d}{2} - m \ell$, $b = \frac{d}{2} + m \ell$, $c = \frac{d}{2} + 1$. This is the same as the $\mathds{H}^d$ solution in \cite{Muck:1999mh}, so we can proceed as follows. We want solutions that decay as a power of $z$ as $z \rightarrow \infty$. There are two independent solutions, which up to overall normalizations $\lambda_\pm$ are
\begin{equation}
\label{eq:gammalambda}
\gamma_{\pm} (z) = \lambda_{\pm} z^{-\left( \frac{d}{2} \pm m \ell \right)} {}_2F_1 \left( \frac{d}{2} \pm m \ell, \pm m \ell, 1 \pm 2m \ell; \frac{1}{z} \right).
\end{equation}
Both are allowed for sufficiently small $|m\ell|$, though large $|m\ell|$ requires the $+$ sign.
The expansion of $\alpha_\pm (s)$ around $s = 0$ is
\begin{equation}
\alpha_\pm (s) \approx \lambda_\pm \left( \frac{s}{2\ell} \right)^{2-d} \frac{\Gamma(1 \pm 2m \ell) \Gamma \left( \frac{d}{2} -1\right)}{\Gamma \left( \frac{d}{2} \pm m \ell\right) \Gamma ( \pm m \ell )}.
\end{equation}
The above expression must match with the Minkowski solution in the limit of vanishing $s$, so the coefficients $\lambda_\pm$ are given by
\begin{equation}
\label{eq:lambda}
\lambda_\pm = \mp  2^{-(d \pm 2 m \ell)} \ell^{1-d} \frac{ \Gamma \left( \frac{d}{2} \pm m \ell \right)}{\pi^{(d-1)/2} \Gamma (\frac{1}{2} \pm m \ell)}.
\end{equation}
The choice of sign in these quantities corresponds to a choice of boundary conditions analogous to that referenced in \eqref{eq:cft}. For simplicity, we will choose the $(+)$ boundary condition for all $p$, as this is allowed for any effective 3D fermion mass.  We do so for all computations in the main text.

\subsection{$\expval{T_{kk}}$ in KKZBO}
We still don't have a closed form expression for $S(x,x')$, as we don't know anything about $\Lambda (x,x')$, but we can still calculate observables such as $\expval{\psi (x) \psi (x')}$, $\expval{J_\mu}$, and $\expval{T_{\mu \nu}}$. We'll only carry out the calculation for our quotient spacetime\footnote{For a similar calculation in AdS$_4$ for timelike geodesics, see \cite{ambrus}}.
The expectation value of the Belinfante stress-energy tensor in terms of the spinor propagator is \cite{freedman2012supergravity}
\begin{equation}
\expval{T_{\mu \nu}} = \frac{i}{2} \lim_{x \rightarrow x'} \textrm{Tr} \left[ \left( \gamma_{( \mu}D_{\nu )} S (x, x') - \overline{D_{\nu'}} S(x,x') \gamma_{( \mu} g^{\nu'}_{\nu )} \right) \Lambda (x',x) \right]
\end{equation}
where a primed index on the covariant derivative denotes a derivative on the second coordinate and action from the left. The parallel propagators become trivial at coincident points, namely $g_\nu^{\nu'} = \delta_\nu^{\nu'}$ and $\Lambda (x,x') = \mathds{I}_2$ at $x = x'$. As shown in section \ref{subsec:Tkk}, the null-null component of the stress-energy tensor in the quotient KKZBO spacetime is
\begin{equation}
\expval{T_{kk}} \equiv k^\mu k^\nu \expval{T_{\mu \nu}} = \frac{i}{2} \sum_{A,B}\Tr \left[ \slashed{k} D_k S(x, J_3x') \tilde j \right] + \Tr \left[ \tilde j \overline{D_{k'}} S(J_3 x', x)\slashed{k}\right],
\end{equation}
where we've suppressed the sum over Kaluza-Klein modes $p$ but kept the sum over three-dimensional fermion representations $A$ and $B$. In order for this stress-energy tensor to be real, we expect that the second trace is related to the first by complex conjugation. Using the fact that
\begin{equation}
\gamma^0 S(x,x')^\dagger \gamma^0 = S(x',x)
\end{equation}
the trace of the Hermitian conjugate of the first term becomes
\begin{align}
\Tr \left[ \left( \slashed{k} D_k S(x, J_3 x') \tilde j \right)^\dagger \right] &= \Tr \left[ \tilde j^\dagger D_{k'}^\dagger S(x, J_3x')^\dagger \slashed{k}^\dagger \right] \nonumber \\
&= \Tr \left[ \tilde j^\dagger D_{k'}^\dagger \gamma^0 S(J_3x', x) \gamma^0 \slashed{k}^\dagger \right] \nonumber \\
&= - \Tr \left[ \tilde j \overline{D_{k'}} S(J_3x', x)  \slashed{k} \right]
\end{align}
where we've also used $\gamma^0 \slashed{k} \gamma^0 = \slashed{k}^\dagger$ and $\gamma^0 \tilde j \gamma^0 = -\tilde j = -\tilde j^\dagger$. As the trace of the Hermitian conjugate is the complex conjugate of the original trace, we have
\begin{equation}
\Tr \left[  \slashed{k} D_k S(x, J_3 x') \tilde j \right] = - \Tr \left[ \tilde j \overline{D_{k'}} S(J_3x', x)  \slashed{k} \right]^*
\end{equation}
as expected. The sum of the two traces can then be represented as the imaginary part of the first, and the stress-energy tensor becomes
\begin{equation}
\expval{T_{kk}} = - \sum_{A,B} \Im{\Tr \left[\tilde j \slashed{k} D_k S(x,J_3x') \right]}.
\end{equation}
Plugging in our ansatz \eqref{eq:ansatz} and substituting \eqref{eq:DLambda}, we obtain
\begin{align}
\expval{T_{kk}} &=  - \sum_{A,B} \Im \big\{ \Tr \left[ \tilde j \slashed{k} k^\mu D_\mu \left( (\alpha(s) + \beta(s) \slashed{n})\Lambda (x, J_3 x) \right) \right] \big\} \nonumber \\
&=  - \left( k^\mu n_\mu \right) \sum_{A,B} \Im{  \Tr \left[ \tilde j \slashed{k} \left( \left( \alpha' + \frac{\alpha}{2}(A+C) \right) + \left( \beta' + \frac{\beta}{2}(C-A)\right)\slashed{n}\right)  \Lambda (x, J_3 x) \right]}.
\end{align}
The $A$ and $B$ representations are distinguished by gamma matrices with opposite sign: $\gamma_A^\mu = -\gamma_B^\mu$. The sum over representations will cancel terms containing $\alpha(s)$, as they involve an odd number of gamma matrices, and introduce a factor of 2 to the $\beta(s)$ terms, as they contain an even number of gamma matrices. The stress-energy tensor is therefore
\begin{align}
\expval{T_{kk}} &= - 2  \left( k^\mu n_\mu \right) \left( \beta' + \frac{\beta}{2}(C-A)\right) \Im{ \Tr \left[ \tilde j \slashed{k} \slashed{n}  \Lambda (x, J_3 x) \right]} \nonumber \\
&= - 2  \left( k^\mu n_\mu \right) \left( \frac{\partial}{\partial s} - \frac{1}{2 \ell} \coth \frac{s}{2 \ell}\right) \beta(s) \Im{\Tr \left[ \tilde j \slashed{k} \slashed{n}  \Lambda (x, J_3 x) \right]}.
\end{align}
In general, calculating $\Lambda(x,x')$ involves a path ordered integral along the geodesic in question, so it's easier to choose coordinates such that the parallel propagator becomes trivial even for noncoincident points. As our calculation ultimately takes place in the covering space AdS$_3$, we can choose global AdS$_3$ coordinates $(t,\rho,\varphi)$ with metric
\begin{equation}
ds^2 = -\left( 1 + \rho^2/\ell^2 \right) dt^2 + \frac{d\rho^2}{1 + \rho^2/\ell^2}  + \rho^2 d \varphi^2.
\end{equation}
For a timelike slice of AdS$_3$, we define the origin by the intersection of the timelike axis defined by the isometry and the geodesic itself as in Figure \ref{fig:rp2}. We therefore have geodesic norm $n_\mu = \frac{1}{\sqrt{1+\rho^2/\ell^2}} \partial_\rho$ and a perpendicular unit vector $\varphi_\mu = \rho \partial_\varphi$. In these coordinates, the spinor parallel propagator for our spacelike geodesic is trivial, as can be seen from the covariance of $\Lambda (x,x')$ along the geodesic:
\begin{align}
n^\mu D_\mu \Lambda(x,x') &= n^\rho \partial_\rho \Lambda (x,x') = 0 \nonumber \\
\Rightarrow \Lambda(x,x') &= \mathds{I}_2.
\end{align}
Additionally, as the vielbein for this coordinate system is diagonal, we can rewrite the isometry $\tilde j = i \gamma^1 \gamma^2$ as
\begin{equation}
\tilde j = i \slashed{n} \slashed{\varphi}.
\end{equation}
The trace inside of the stress-energy tensor therefore becomes
\begin{align}
\Tr \left[ \tilde j \slashed{k} \slashed{n}\right] &= \Tr \left[i \slashed{k} \slashed{\varphi} \right] \nonumber \\
&= 2 i k^\mu \varphi_\mu.
\end{align}
This expression is manifestly vielbein independent. The final expression for the null-null component of the stress-energy tensor is therefore
\begin{equation}
\expval{T_{kk}} = -4 \left( k^\mu n_\mu\right) \left( k^\mu \varphi_\mu \right) \left( \frac{\partial}{\partial s} - \frac{1}{2 \ell} \coth \frac{s}{2 \ell}\right) \beta(s).
\end{equation}
The full calculation involves a sum over Kaluza-Klein modes $p$ and BTZ images $n$. Explicitly, in terms of our Kruskal-like coordinates $(U,V,\phi)$ and $\rho = \ell \sqrt{\sinh^2 \left( \frac{r_+ \phi}{\ell}\right) + U^2 \exp \left( -\frac{2 r_- \phi}{\ell} \right)}$, we have
\begin{align}
k^\mu n_\mu &= n_U = \frac{e^{-2 r_- \phi/\ell} \ell^3 U}{\rho \sqrt{\rho^2 + \ell^2}} \nonumber \\
k^\mu \varphi_\mu &= \varphi_U = \frac{e^{- r_- \phi/\ell} \ell^2 \sinh \left(  \frac{r_+ \phi}{\ell}\right)}{\rho}.
\end{align}
\pagebreak
\bibliographystyle{utcaps}
\bibliography{spinorWormholes}

\end{document}